\documentclass[aps,pre,showpacs,amsmath,amssymb,twocolumn]{revtex4}\usepackage{pslatex}
\usepackage{graphicx}
\usepackage{dcolumn}
\usepackage{natbib}

\begin{document}

\title{Logarithmic Relaxation in a Colloidal System}

\author{M. Sperl}
\affiliation{Physik-Department, Technische Universit\"at M\"unchen, 
85747 Garching, Germany}

\date{\today}

\begin{abstract}
The slow dynamics for a colloidal suspension of particles interacting with
a hard-core repulsion complemented by a short-ranged attraction is
discussed within the frame of mode-coupling theory for ideal glass
transitions for parameter points near a higher-order glass-transition
singularity. The solutions of the equations of motion for the density
correlation functions are solved for the square-well system in
quantitative detail by asymptotic expansion using the distance of the
three control parameters packing fraction, attraction strength and
attraction range from their critical values as small parameters. For given
wave vectors, distinguished surfaces in parameter space are identified
where the next-to-leading order contributions for the expansion vanish so
that the decay functions exhibit a logarithmic decay over large time
intervals. For both coherent and tagged particle dynamics the
leading-order logarithmic decay is accessible in the liquid regime for
wave vectors of several times the principal peak in the structure factor.
The logarithmic decay in the correlation function is manifested in the
mean-squared displacement as a subdiffusive power law with an exponent
varying sensitively with the control parameters. Shifting parameters
through the distinguished surfaces, the correlation functions and the
logarithm of the mean-squared displacement considered as functions of the
logarithm of the time exhibit a crossover from concave to convex behavior,
and a similar scenario is obtained when varying the wave vector.
\end{abstract}


\maketitle

\section{\label{sec:introduction}Introduction}

The dynamics in an interacting many particle system is conveniently
described by density autocorrelation functions $\phi_q(t)$ for time $t$
and wave vector $q$. These correlation functions can be measured in both
experiment and computer simulation \cite{Boon1980}.  Mode-coupling theory
for ideal glass transitions (MCT) discusses the transition from a liquid
to a glass as a bifurcation in the long-time limit of the correlator
$\phi_q(t)$ \cite{Goetze1991b}. In the liquid state, the correlation
function decays to zero. If a control parameter, say density, exceeds some
critical value, the long-time limit changes discontinuously from zero to a
finite value, a glass transition occurs \cite{Bengtzelius1984}. This
liquid-glass transition is identified with an $A_2$- or fold singularity
\cite{Arnold1986} in the equations of motion of MCT. The simplest example
for a liquid-glass transition is found in the hard-sphere system (HSS),
where the interaction potential among the particles is zero unless their
mutual separation becomes smaller than their diameter where the potential
becomes infinitely repulsive, thus preventing the particles from
overlapping. The HSS is the system MCT was applied to first
\cite{Bengtzelius1984}, and it is also the system for which the most
detailed predictions have been worked out \cite{Franosch1997,Fuchs1998}.
Close to the singularity, the equations of motion can be expanded in
asymptotic series. This yields a two-step decay with two related power
laws for the short-time and the long-time decay at a liquid-glass
transition \cite{Goetze1991b}. The HSS can be realized in colloidal
suspensions \cite{Pusey1986}. Experiments in these systems lead to the
conclusion that MCT is able to describe the main aspects of the glass
transition qualitatively and some aspects even quantitatively
\cite{Megen1995}.

MCT can also exhibit other singularities than the fold \cite{Goetze1991b}.  
These higher-order singularities were predicted recently to occur for
colloidal systems with short-ranged attraction where $A_3$- and
$A_4$-singularities were found that are also called cusp and swallowtail
\cite{Fabbian1999,Bergenholtz1999,Dawson2001}. In these systems, the
hard-core repulsion is supplemented by a short ranged attraction, e.g. in
the square-well system (SWS). A cusp singularity is the endpoint of a line
of glass-glass transitions that arises if two different mechanisms of
arrest are of the same importance. In the SWS the first mechanism is the
hard core repulsion that causes a transition as in the HSS via the well
known cage effect. The second mechanism leading to arrest is bond
formation introduced by the attractive part of the potential.  This latter
transition was proposed as relevant for the transition to a gel
\cite{Bergenholtz1999}. If the difference in the two mechanisms is less
pronounced, the glass-glass transitions vanish and give rise to an
$A_4$-singularity. In the SWS this happens as the range of the attraction
is increased \cite{Dawson2001}. The range of attraction considered here is
of order less than $20\%$ of the particle diameter and the strength is
about several $k_\text{B}T$. This is accessible in colloid-polymer
mixtures with nonadsorbing polymer which is well under control
experimentally \cite{Poon2002}. Higher-order singularities have also been
identified for a number of short-ranged potentials with shapes differring from 
the SWS yielding certain quantitative trends but no qualitative changes
\cite{Goetze2003b}.

In addition to the success of MCT for the description of the HSS, two
findings support the use of this theory for the description of colloids
with attraction. First, a reentry phenomenon was predicted by the theory
where a glass state is melted upon increasing the attraction
\cite{Fabbian1999,Dawson2001}. This was subsequently found in several
experiments \cite{Eckert2002,Pham2002} and computer simulation studies
\cite{Pham2002,Foffi2002b,Zaccarelli2002b,Puertas2003}. Second, there are
indications of logarithmic decay \cite{Puertas2002} and related anomalous
decays \cite{Mallamace2000,Zaccarelli2002b} that are consistent with
scenarios found within MCT \cite{Fabbian1999,Dawson2001}. To investigate
the dynamics in such systems, apart from computer simulation dynamic light
scattering has already been used to determine correlation functions
\cite{Mallamace2000,Eckert2002,Pham2002}. Direct imaging techniques are
available to determine also the mean-squared displacement (MSD) with high
precision \cite{Kegel2000,Weeks2000,Weeks2002}. The purpose of the present
paper is the application of the general theory for higher-order
singularities \cite{Goetze2002} to the SWS and the derivation of testable
quantitative predictions for the correlation functions and the MSD.
Certain scenarios have been discussed before for schematic models 
\cite{Goetze2002}. Some of these scenarios shall be identified also in the 
microscopic model in the following.

The paper is organized as follows. In Sec.~\ref{sec:asy} the equations of
motion and the asymptotic solution for the logarithmic decay are
summarized and the subdiffusive power law for the MSD is derived. The
theory is applied to the $A_4$-singularity in the SWS for the correlation
functions in Sec.~\ref{sec:corr} and for the MSD in Sec.~\ref{sec:MSD}.
Changes in the scenarios when moving from the $A_4$-singularity to
$A_3$-singularities are discussed in Sec.~\ref{sec:A3}, and
Sec.~\ref{sec:HCY} contains a comparison to results obtained for the
hard-core Yukawa system (HCY).  Section\ref{sec:conclusion} presents a
conclusion.

\section{\label{sec:asy}Asymptotic Solutions}

We shall consider a system of $N$ particles with diameter $d$ in a volume
$V$ interacting with a spherical potential. When at time $t$ the $j$th
particle is located at $\vec{r}_j(t)$ the density variables are defined as
$\rho_q(t)=\sum_j\exp[i\vec{q}\vec{r}_j(t)]$.

\subsection{\label{subsec:eom}Equations of Motion}

The equations of motion for the normalized density correlators $\phi_q(t)=
\langle\rho_{{\vec{q}}}^{*}(t)\rho_{\vec{q}}\rangle/\langle|\rho_{\vec{q}}|^2
\rangle$ within MCT, when Brownian dynamics for the motion in colloids is
assumed, are given by
\cite{Bengtzelius1984,Goetze1991b,Szamel1991,Fuchs1993a},
\begin{subequations}\label{eq:mct:MCT}
\begin{equation}\label{eq:mct:phi_col}
\tau_q\partial_t\phi_q(t)+\phi_q(t)+\int_0^t
m_q(t-t')\partial_{t'}\phi_q(t')\,dt'=0\,.
\end{equation}
The initial conditions are $\phi_q (0)=1$. The microscopic time scale
reads $\tau_q=S_q/(D_0q^2)$. It is given by the short-time diffusion
coefficient, $D_0$, characterizing the Brownian motion and the static
structure factor $S_q=\langle|\rho_{\vec{q}}|^2\rangle$. The mode-coupling
approximation results in expressing the kernels $m_q(t)$ in terms of the
correlators $\phi_q(t)$ \cite{Goetze1991b},
\begin{equation}\label{eq:mct:kernel}
m_q (t) = {\cal F}_q \left[\mathbf{V}, \phi_k (t) \right] \,\, .
\end{equation}
As a consequence of the factorization into pair modes for the structural
relaxation in simple liquids, ${\mathcal F}_q$ is a bilinear functional of
the density correlators \cite{Bengtzelius1984},
\begin{equation}\label{eq:mct:Fdef}
{\mathcal F}_q[\tilde{f}]  = \frac{1}{2} \int \frac{{d}^3k}{(2 \pi )^3}
V_{\vec{q},\vec{k}} \tilde{f}_k \tilde{f}_{|\vec{q}-\vec{k}|}\,,
\end{equation}
and the vertex is determined completely by the static structure of the
liquid system \cite{Goetze1976b,Sjoegren1980b},
\begin{equation}\label{eq:mct:vertex}
V_{\vec{q},\vec{k}} =  S_q S_k S_{|\vec{q}-\vec{k}|}\, \rho
\left[ {\vec{q}} \cdot
\vec{k}\,{c_k} +\vec{q} \cdot
(\vec{q}-\vec{k})\,{{c_{|\vec{q}-\vec{k}|}} }
 \right]^2/q^4\,.
\end{equation}
\end{subequations}
The number density is given by $\rho=N/V$ and $c_q$ denotes the direct
correlation function which is related to the static structure factor $S_q$
in the Ornstein-Zernike relation, $S_q=1/[1-\rho\,c_q]$, both depend on
external control parameters like density or temperature \cite{Hansen1986}.
For the SWS with hard-core diameter $d$, depth of the potential $u_0$, and
range of the potential $\Delta$, we get three dimensionless control
parameters, the packing fraction $\varphi=d^3\rho\pi/6$, the attraction
strength $\Gamma=u_0/(k_\text{B}T)$ and the relative well width
$\delta=\Delta/d$. These can be combined to a control-parameter vector
$\mathbf{V}=(\varphi,\Gamma,\delta)$.

It is the long-time limit of the correlation function,
$\lim_{t\rightarrow\infty}\phi_q(t) = f_q$, that determines whether a
system is in the liquid regime, where $f_q=0$, or in an arrested state,
where $0< f_q\leqslant 1$. In the latter case, the values $f_q$
characterize the arrested glassy state and the $f_q$ are called glass-form
factors or Debye-Waller factors. In the long-time limit, the equation of
motion Eq.\ (\ref{eq:mct:MCT}) reduces to an equation involving only the
mode-coupling functional and the glass-form factors \cite{Goetze1991b},
\begin{equation}\label{eq:Feq}
f_q/(1-f_q) = {\cal F}_q[f]\,.
\end{equation}

Frequently studied is the dynamics of a single or tagged particle with the
single particle density $\rho_q^s(t)=\exp[i\vec{q}\vec{r}_s(t)]$. For the
correlation function of a tagged particle,
$\phi^s_q(t)=\langle\rho_{{\vec{q}}}^{s\,*}(t) \rho^s_{\vec{q}}\rangle$,
similar equations as Eqs.\ (\ref{eq:mct:MCT}) have been derived
\cite{Bengtzelius1984,Fuchs1998},
\begin{subequations}\label{eq:mct:tagged}
\begin{equation}\label{eq:mct:phis_col}
\tau^s_q\partial_t\phi^s_q(t)+\phi^s_q(t)+\int_0^t
m^s_q(t-t')\partial_{t'}\phi^s_q(t')\,dt'=0\,,
\end{equation}
with $\tau^s_q=1/(D^s_0q^2)$. The short-time diffusion coefficient for a
single particle, $D^s_0$, again specifies the Brownian dynamics. The
mode-coupling functional for the tagged particle motion,
\begin{equation}\label{eq:mct:Fsdef}
{\mathcal F}^s_q[f,f^s]  = \int \frac{{d}^3k}{(2 \pi )^3}  S_k
\frac{\rho}{q^4} {c^s_k}^2  (\vec{q}\vec{k})^2
f_k f^s_{|\vec{q}-\vec{k}|}\,,
\end{equation}
is also determined by the static structure of the liquid system where
$c^s_q$ is the single-particle direct correlation function
\cite{Hansen1986}.
\end{subequations}

The dynamics of the tagged particle is coupled to the coherent density
correlator $\phi_q(t)$ and for that reason $\phi^s_q(t)$ also displays the
bifurcation dynamics that is driven by $\phi_q(t)$. The equation for the
long-time limits of the tagged particle correlations function,
$\phi^s_q(t\rightarrow\infty)=f_q^s$, reads
\begin{equation}\label{eq:mct:fqs}
f_q^s/(1-f_q^s)
={\cal F}^s_q[f,f^s]\,.
\end{equation}
In the following, the tagged particle will be assumed as of the same sort
as the host fluid. If the host particles are in the liquid state, $f_q=0$,
a tagged particle cannot be arrested, and in that case Eq.\
(\ref{eq:mct:fqs})  implies $f^s_q=0$.

The MSD is defined by 
$\delta r^2(t) = \langle |\vec{r}_s(t)-\vec{r}_s(0)|^2 \rangle$ and 
describes the average distance a particle has traveled within some time
$t$ \cite{Hansen1986}. It is obtained, e.g., as small wave-number limit of
the tagged-particle correlator in Eq.\ (\ref{eq:mct:tagged}),
$\phi^s_q(t)=1-q^2\delta r^2(t)/6+{\mathcal O}(q^4)$
\cite{Goetze1991b,Fuchs1998},
\begin{subequations}\label{eq:mct:MSD}
\begin{equation}\label{eq:mct:MSD_col}
\delta r^2(t) + D^s_0 \int_0^t m^{(0)}(t-t')\,\delta
r^2(t')\,dt'=6D^s_0t\,,
\end{equation}
$m^{(0)}(t)=\lim_{q\rightarrow 0}m^s_q(t) 
={\mathcal F}_{MSD}[\phi(t),\phi^s(t)]$. The mode-coupling functional for 
the MSD reads
\begin{equation}\label{eq:mct:FMSDdef}
{\mathcal F}_{MSD}[f,f^s] = \int \frac{{d}k}{(6 \pi^2 )}\,\rho\, S_k
(c^s_k)^2 f_k f^s_k\,.
\end{equation}
\end{subequations}
A characteristic localization length $r_s$ is defined by the second moment
for the relaxation of the distribution of $\phi^s_q(t)$
\cite{Goetze1991b}, which can be identified with the functional in Eq.\
(\ref{eq:mct:FMSDdef}) $r_s^2=1/{\mathcal F}_{MSD}[f,f^s]$. It is the
long-time limit of the MSD. Its value at the critical point, $r_s^c$,
characterizes the arrested structure. The value $6r_s^{c\,2}$ represents
the plateau for the dynamics of $\delta r^2(t)$.

Equations~(\ref{eq:mct:MCT}) to (\ref{eq:mct:MSD}) are solved numerically
using algorithms introduced in \cite{Fuchs1991b,Goetze1996}. Details of
the implementation are found in \cite{Franosch1997,Dawson2001}. We use $d$
as unit of length, $d=1$, and choose the unit of time so that
$1/D_0=1/D_0^s=160$. The structure factors for the SWS and the HCY are
calculated in mean-spherical approximation \cite{Dawson2001,Cummings1979}.
The wave numbers shall be discretized to a grid of $M$ points with a
spacing $\Delta q=0.4/d$. The cutoff in the calculations is ranging from
$M=300$ for $\delta>0.04$ up to $M=750$ for $\delta=0.02$.

\subsection{\label{subsec:log}Logarithmic Decay Laws}

The asymptotic solution at higher-order singularities shall be quoted from
Ref.~\cite{Goetze2002} where also further details can be found. The
asymptotic expansion is performed in small deviations of the correlation
function from the critical long-time limit $f^c_q$ involving the
coefficients
\begin{equation}\label{eq:asy:Aqk_def}
A_{q k_1 \cdots k_n}^{(n)} (\mathbf{V})  =  \frac{1}{n!} (1 - f_q^c)\,
\frac{\partial^n {\cal F}_q \left [\mathbf{V}, f_k^c \right ] }{ \partial
f_{k_1}^c \cdots\partial f_{k_n}^c } \,
(1 - f_{k_1}^c) \cdots (1 - f_{k_n}^c)   \,,
\end{equation}

which can be split into values at the singularity, 
$A_{qk_1\cdots k_n}^{(n)c}$, and remainders, 
$A_{q k_1 \cdots k_n}^{(n) } (\mathbf{V}) = A_{q k_1 \cdots k_n}^{(n)c} 
+ \hat{A}_{q k_1 \cdots k_n}^{(n)} (\mathbf{V})$.
The Jacobian matrix of Eq.\ (\ref{eq:Feq}) is singular at the critical 
points and assumes the form $\bigl[\delta_{q k} - A_{q k}^{(1) c} \bigr]$. 
The non-negative left and right eigenvectors of matrix $A_{q k}^{(1) c}$ 
shall be denoted by $a^*_q$ and $a_q$ and can be fixed uniquely by 
requiring $\sum_q \, a^*_q \, a_q = 1$ and $\sum_q \, a^*_q \, a_q^2 = 1$. 
The reduced resolvent $R_{q k}$ of $A_{q k}^{(1) c}$ maps vectors 
orthogonal to $a^*_q$ to vectors orthogonal to $a_q$.
The leading-order result for Eq.\ (\ref{eq:mct:MCT}) is then given by
\begin{equation}\label{eq:asy:log_decay}
\phi_q (t) = f_q^c + h_q \left[ - B \ln (t / \tau) \right]  \,,\;
 B = \sqrt{\left[ - 6\varepsilon_1 (\mathbf{V}) / \pi^2 \right]}\,,
\end{equation}
with the critical amplitudes $h_q=(1-f^c_q)a_q$ and the separation 
parameter $\varepsilon_1(\mathbf{V})=a^*_q \hat{A}_q^{(0)}(\mathbf{V})$, 
which is restricted to negative values, $\varepsilon_1<0$. If 
$\varepsilon$ indicates the distance of the control parameters 
$\mathbf{V}$ from the critical point, the leading result is of order 
$\sqrt{\varepsilon}$ and correct up to ${\cal O}(\varepsilon)$. The 
next-to-leading-order approximation is
\begin{eqnarray}\label{eq:asy:G1G2q}
\phi_q (t) = (f_q^c + \hat{f}_q) &+& h_q  \bigl [  (- B + B_1)
\ln (t / \tau)  \nonumber
\\ && +  (B_2 + K_q B^2) \ln^2 (t / \tau)\nonumber
\\ && +     B_3 \ln^3 (t / \tau) +  B_4 \ln^4 (t / \tau)  \bigr ] \,,
\end{eqnarray}
what includes the terms of order $\varepsilon$ and neglects terms of order
$\varepsilon^{3/2}$. It involves corrections to the plateau values,
\begin{equation}\label{eq:asy:deltafq}
\hat{f}_q = (1 - f_q^c)  R_{q k} \left[ \hat{A}_{k }^{(0)} (\mathbf{V}) -
\epsilon_1 (\mathbf{V}) a_k^2 \right]\,,
\end{equation}
correction amplitudes,
\begin{equation}\label{eq:asy:Kq}
K_q =  R_{q k} \left[ A_{k k_1 k_2 }^{(2)c} a_{k_1} a_{k_2} 
- a_{k}^2 \right] / a_q  \,,
\end{equation}
and the prefactors,
\begin{subequations}\label{eq:asy:Bcoeffs}
\begin{equation}\label{eq:asy:Bcoeffs_B1}
B_1 =  (0.44425\, \zeta - 0.065381\, \mu_3)\,\varepsilon_1 (\mathbf{V})
- 0.22213 \,\varepsilon_2 (\mathbf{V}) \,\, ,
\end{equation}
\begin{equation}\label{eq:asy:Bcoeffs_B2}
B_2 = (0.91189\,\zeta + 0.068713\,\mu_3)\,\varepsilon_1 (\mathbf{V})
- 0.15198\,\varepsilon_2 (\mathbf{V}) \,\, ,
\end{equation}
\begin{equation}\label{eq:asy:Bcoeffs_B3}
B_3 = - 0.13504 \,\mu_3 \,\varepsilon_1 (\mathbf{V}) \,\, ,
\quad B_4 = - 0.046197 \,\mu_3 \,\varepsilon_1 (\mathbf{V}) \,\, .
\end{equation}
\end{subequations}
The numbers characterizing the higher-order singularities are
\begin{equation}\label{eq:asy:zeta}
\zeta = \sum_q a^*_q \left[ a_q^2 K_q +  a_q^3/2
\right] \,\, ,
\end{equation}
and
\begin{equation}\label{eq:asy:mu3}
\mu_3 = 2 \zeta - \sum_q a^*_q \left[ A_{q k_1 k_2 k_3}^{(3)c}
a_{k_1} a_{k_2} a_{k_3} + 2  A_{q k_1 k_2}^{(2)c} a_{k_1}a_{k_2}
K_{k_2}
 \right] \,\, .
\end{equation}
For the leading correction also an additional separation parameter is 
introduced,
\begin{equation}\label{eq:asy:epsilon2_A}\begin{split}
\varepsilon_2 (\mathbf{V}) =  \sum_q a^*_q
        \hat{A}_{q k}^{(1)} (\mathbf{V}) a_k
+ 2 \varepsilon_1 (\mathbf{V}) \sum_q a^*_q a_q^2 K_q 
\\  +  2 \sum_q a^*_q \left[ A_{q k_1 k_2}^{(2)c} a_{k_1}
\hat{f}_{k_2}/(1-f_{k_2}^c) 
- a_q \hat{f}_q /(1-f_q^c) \right]  \,.\end{split}
\end{equation}
The time scale $\tau$ is determined by matching asymptotic approximation 
and numerical solution of $\phi_q(t)$ at the plateau $f_q^c$ or the 
rescaled plateau $f_q^c+\hat{f}_q$ for Eq.\ (\ref{eq:asy:log_decay}) and 
Eq.\ (\ref{eq:asy:G1G2q}), respectively.

\subsection{\label{subsec:couple}Coupled Variables}

Inserting the asymptotic expansion of Eq.\ (\ref{eq:asy:G1G2q}) into the
long-time limit of Eq.\ (\ref{eq:mct:phis_col}), the approximation for the
tagged particle dynamics up to order $\varepsilon$ is obtained
\cite{Sperl2003},
\begin{eqnarray}\label{eq:asy:phis}
\phi^s_q (t) = (f_q^{s\,c} + \hat{f}^s_q) &+& h^s_q  \bigl [  (- B + B_1)
\ln (t / \tau)  \nonumber
\\ && +  (B_2 + K^s_q B^2) \ln^2 (t / \tau)\nonumber
\\ && +     B_3 \ln^3 (t / \tau) +  B_4 \ln^4 (t / \tau)  \bigr ]\,,
\end{eqnarray}
with the critical amplitudes $h^s_q=(1-f^{s\,c}_q)\,a_q^s$, the correction
amplitudes $K^s_q$ and the plateau corrections $\hat{f}^s_q$. The latter 
are derived from the functional~(\ref{eq:mct:Fsdef}) and the related 
coherent quantities by
\begin{subequations}\label{eq:asy:taggedTaylor}
\begin{equation}\label{eq:asy:taggedTaylor:asq}
\sum_k (\delta_{qk} - {A^{s\,c}_{q,k}})\, a_k^s = \sum_k  
{A^{s\,c}_{qk}} a_k\,,
\end{equation}
\begin{equation}\label{eq:asy:taggedTaylor:Ksq}\begin{split}
\sum_k (\delta_{qk} - {A^{s\,c}_{q,k}})\,a_k^s K^s_k = -
{a^s_q}^2
+\sum_k {A^{s\,c}_{qk}} a_k K_k\\
+\sum_{k,p} [
{A^{s\,c}_{q,kp}} a^s_k a^s_p + {A^{s\,c}_{qkp}} a_k a_p + 
{A^{s\,c}_{qk,p}} a_k a^s_p
]\,,\end{split}
\end{equation}
\begin{equation}\label{eq:asy:taggedTaylor:fsq}
\sum_k (\delta_{qk} - {A^{s\,c}_{q,k}})\,a_k^s \hat{f}_k^s = 
-\varepsilon_1(\mathbf{V})\,{a^s_q}^2
+ \sum_k {A^{s\,c}_{qk}}\, a_k\hat{f}_k
+\hat{A}^{s}_q(\mathbf{V})
\,.
\end{equation}
\end{subequations}
The derivatives with respect to the coherent and tagged particle 
glass-form factors  are denoted before and after the comma, respectively. 
The coefficients are 
\begin{equation}\label{eq:asy:Asqk_def}\begin{split}
&A^s_{q k_1 \cdots k_n, p_1 \cdots p_m} (\mathbf{V})  =  
\frac{1}{n!} \frac{1}{m!} (1 - {f^s_q}^c)\,
\frac{\partial^n \partial_m {\cal F}^s_q \left [\mathbf{V}, f_k^c 
,{f^s_q}^c\right ] }{ \partial f_{k_1} \cdots\partial f_{k_n}
\partial f^s_{p_1} \cdots\partial f^s_{p_n} } \,
\\&\qquad\times
(1 - f_{k_1}^c) \cdots (1 - f_{k_n}^c)  
(1 - {f^s_{p_1}}^c) \cdots (1 - {f^s_{p_n}}^c) =\\
&\qquad={A^{sc}_{q k_1 \cdots k_n, p_1 \cdots p_m}} +
\hat{A}^{s}_{q k_1 \cdots k_n, p_1 \cdots p_m} (\mathbf{V})
\,\, .\end{split}
\end{equation}

Similar arguments as above yield the asymptotic expansion for the MSD up 
to order $\varepsilon$ ,
\begin{eqnarray}\label{eq:asy:MSD}
\frac{1}{6}\delta r^2 (t) = {r^c_{s}}\,^2 &-& {\hat{r}_{s}}^2 -  
h_\text{MSD}\, \bigl [  (- B + B_1)
\ln (t / \tau)  \nonumber
\\ && +  (B_2 + K_\text{MSD} B^2) \ln^2 (t / \tau)\nonumber
\\ && +     B_3 \ln^3 (t / \tau) +  B_4 \ln^4 (t / \tau)  \bigr ] \,,
\end{eqnarray}
with parameters
\begin{subequations}\label{eq:asy:MSDpar}
\begin{equation}\label{eq:asy:MSDpar:h}
h_\text{MSD} = {r_s^{c\,4}}\{{\cal F}^c_\text{MSD}[h_k,f_p^{s\,c}]+
{\cal F}^c_\text{MSD}[f_k^c,h_p^s]\}\,,
\end{equation}
\begin{equation}\label{eq:asy:MSDpar:K}\begin{split}
K_\text{MSD} = \,&r_s^{c\,4}\{{\cal F}^c_\text{MSD}[h_k,h_p^s]
+{\cal F}^c_\text{MSD}[h_k K_k,f_p^{s\,c}]
\\&\qquad+{\cal F}^c_\text{MSD}[f_k^c,h_p^s K_p^s]
\}/h_\text{MSD}
- 
h_\text{MSD}/r_s^{c\,2}\,,
\end{split}\end{equation}
\begin{equation}\label{eq:asy:MSDpar:df}\begin{split}
\hat{r}^2_{s} =\,& {r_s^{c\,4}}\{
{\cal F}^c_\text{MSD}[h_k\hat{f}_k,f_p^{s\,c}]
+{\cal F}^c_\text{MSD}[f_k^c,h_p^s\hat{f}_p^s]\\&\qquad
+{\cal F}^c_\text{MSD}[f_k^c,f_p^{s\,c}](\mathbf{V})
-{\cal F}^c_\text{MSD}[f_k^c,f_p^{s\,c}](\mathbf{V}^c)
\}/h_\text{MSD}\\&
-\varepsilon_1(\mathbf{V})h^2_\text{MSD}/r_s^{c\,2}\,.
\end{split}\end{equation}
\end{subequations}
For the generic liquid-glass transition, the asymptotic expansion was
carried out with a different convention as in Eq.\ (\ref{eq:asy:Aqk_def}),
however, the quantities $h_q$, $K_q$, $h^s_q$, $K^s_q$, $h_\text{MSD}$,
$K_\text{MSD}$, and $\zeta$ are the same in both descriptions
\cite{Franosch1997,Fuchs1998}.  The plateau corrections, $\hat{f}_q$,
$\hat{f}^s_q$, and $\hat{r}^2_{s}$ are different for liquid-glass
transitions and higher-order singularities. The expansions in Eqs.\
(\ref{eq:asy:G1G2q}), (\ref{eq:asy:phis}), and (\ref{eq:asy:MSD}) share
the coefficients $B$, $B_1$, $B_2$, $B_3$, and $B_4$. They differ in the
plateau and its correction, the critical amplitude and the correction
amplitude.

\subsection{\label{subsec:MSD}Subdiffusive Power Law in the MSD}

The logarithmic decay laws shall be phrased for the MSD in a slightly
different form than in Eq.\ (\ref{eq:asy:MSD}). This is done in order to
account for the fact that the MSD is conveniently shown in a
double-logarithmic representation which is more sensitive to the detection
of power laws. The asymptotic approximation (\ref{eq:asy:MSD}) for the MSD
can be written as $z = a_0 + a_1\,y + a_2\,y^2+ a_3\,y^3 + a_4\, y^4$. 
Here, $z=\delta r^2(t)/6$ and $y=\ln(t/\tau)$ The constant term represents
the square of the corrected localization length,
$a_0=r_s^{c\,2}-\hat{r}_s^2$, the coefficients $a_1=h_\text{MSD}(B-B_1)$,
$a_2=-h_\text{MSD}(B_2+K_\text{MSD}B^2)$ as well as $a_3=-h_\text{MSD}B_3$
and $a_4-h_\text{MSD}B_4$ are the separation dependent prefactors for the
leading and next-to-leading order terms. This yields the expansion
\begin{subequations}\label{eq:asy:expox}
\begin{equation}\label{eq:asy:expoxexp}
\ln z = \ln r_s^{c\,2} - \hat{r}_s^2/r_s^{c\,2} 
+ x' \,y + b_2\,y^2 + a_3/r_s^{c\,2}\, y^3 + a_4/r_s^{c\,2}\, y^4
+ {\mathcal O}(\varepsilon^{3/2})\,,
\end{equation}
with
\begin{equation}\label{eq:asy:expoxpar}
x' = a_1/r_s^{c\,2}\,,\quad b_2= 
\frac{2 r_s^{c\,2} a_2 - a_1^2}{2 r_s^{c\,4}}\,.
\end{equation}
\end{subequations}
In leading order, one gets a power law for the MSD,
\begin{subequations}\label{eq:asy:powerlead}
\begin{equation}\label{eq:asy:expoxlaw}
\delta r^2(t)/6= r_s^{c\,2}\,(t/\tau)^{x}\,,
\end{equation}
with an exponent
\begin{equation}\label{eq:asy:expoxcalc}
x = h_\text{MSD}B/{r_s^c}\,^2\,.
\end{equation}
\end{subequations}
Exponent $x$ varies with the square-root in the separation parameter
$\varepsilon_1$, cf. Eq.\ (\ref{eq:asy:log_decay}). Including the
corrections of order $\varepsilon$ rescales the exponent to
\begin{subequations}\label{eq:asy:powercorr}
\begin{equation}\label{eq:asy:b1dash}
x' = h_\text{MSD}(B-B_1)/r_s^{c\,2}\,.
\end{equation}
and the next-to-leading order result reads
\begin{equation}\label{eq:asy:powercorrb2}\begin{split}
\delta r^2(t)/6 = (t/\tau)^{x'}&\{r_s^{c\,2}\,
-\hat{r}_s^2+ b_2\,r_s^{c\,2} \ln(t/\tau)^2\\&
+a_3\ln(t/\tau)^3+a_4\ln(t/\tau)^4\}\,.\end{split}
\end{equation}
\end{subequations}

\section{\label{sec:corr}Correlation Functions near an $A_4$-singularity}

Before we can apply the asymptotic expansion of Eq.\ (\ref{eq:asy:G1G2q}),
we need to specify the values for $\mu_3$ and $\zeta$ appearing in the
prefactors of Eqs.~(\ref{eq:asy:Bcoeffs}). The $A_4$-singularity is
characterized by $\mu_3=0$. This condition has been used to locate the
$A_4$-singularity at $\mathbf{V}=\mathbf{V}^*$ for the SWS by:
\begin{equation}\label{eq:gtd:A4.SWS.MSA2nd}
\varphi^* = 0.52768\,,\quad\Gamma^* = 4.4759\,,\quad \delta^* = 0.04381\,.
\end{equation}
The vanishing parameter $\mu_3$ implies a considerable simplification in
the preceding formulas since $B_3=B_4=0$ \cite{Goetze2002}. The deviations
of the control parameter values specifying the $A_4$-singularity from the
ones reported in Ref.~\cite{Dawson2001} originate from refined numerical
procedures used here and they do not exceed $6\%$. The characteristic
parameter was $|\mu_3|<5\cdot 10^{-4}$ at the control-parameter values
specified above. The parameter $\zeta$ varies regularly and is
$\zeta=0.122$ at the $A_4$-singularity. This is smaller than the value in
the HSS, $\zeta_\text{HSS}=0.269$ \cite{Franosch1997}.

\begin{figure}[htb] 
\includegraphics[width=\columnwidth]{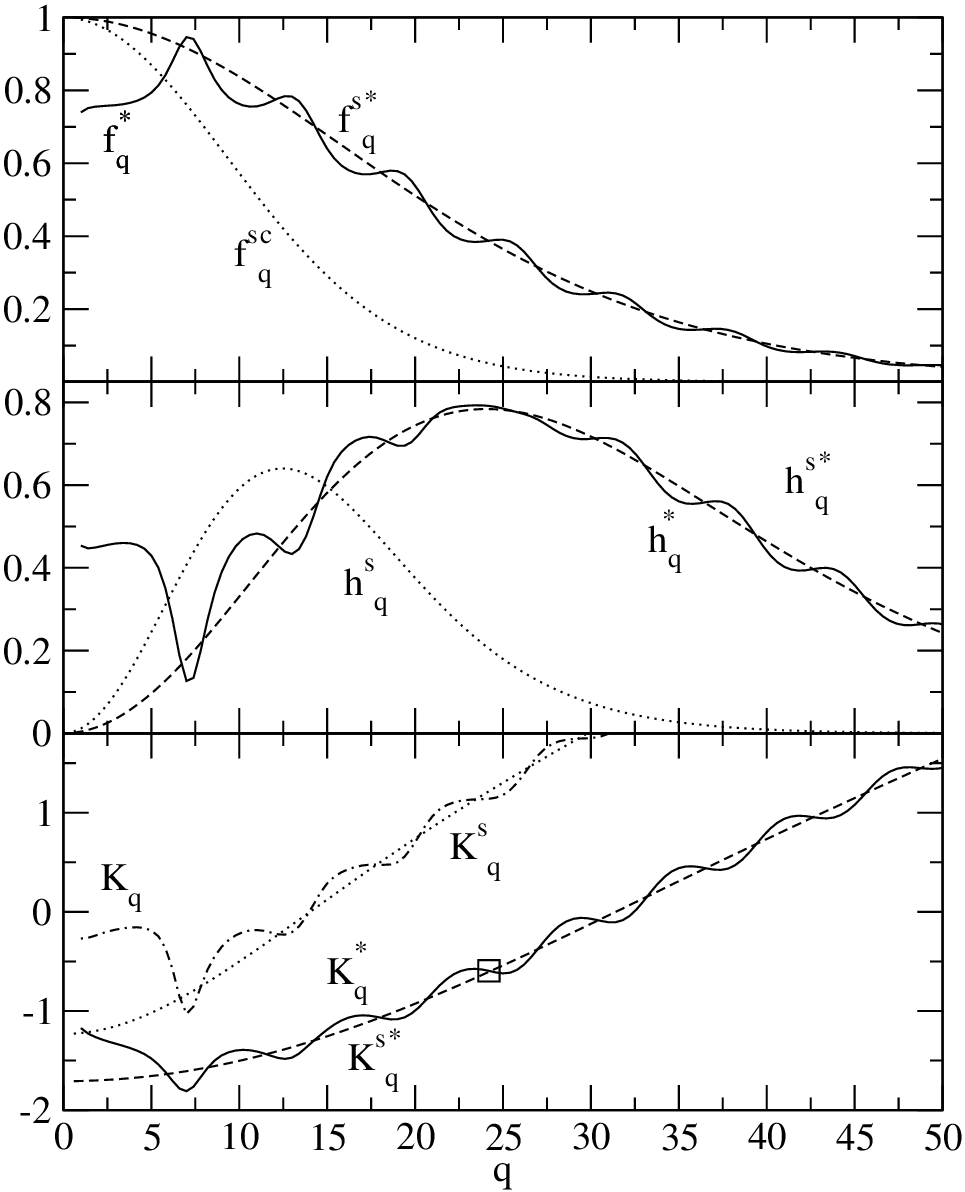}
\caption{\label{fig:swsasy:fqA4}Wave-vector dependent amplitudes 
characterizing the $A_4$-singularity, Eq.\ (\ref{eq:gtd:A4.SWS.MSA2nd}), 
for coherent and tagged particle correlators of the square-well system 
(SWS).
In the upper and middle panel the critical glass-form factors $f_q^*$,
Eq.\ (\ref{eq:Feq}), and the amplitudes $h_q^*$ are shown as full lines,
respectively. The dashed lines represent the values for $f_q^{s\,*}$, Eq.\
(\ref{eq:mct:fqs}), and $h_q^{s\,*}$, Eq.\
(\ref{eq:asy:taggedTaylor:asq}). For the hard-sphere system (HSS),
$f_q^{s\,c}$ and $h_q^{s}$ are shown dotted. The lower panel shows the
correction amplitudes $K_q^*$, Eq.\ (\ref{eq:asy:Kq}), and $K_q^{s\,*}$,
Eq.\ (\ref{eq:asy:taggedTaylor:Ksq}), as full and dashed lines,
respectively.  A square at $q=24.2$ indicates the corrections for the path
calculated for Fig.\ \ref{fig:swsasy:A4quadlines} and the correlators
shown in Fig.\ \ref{fig:swsasy:A4log}. The correction amplitudes $K_q$
($-\cdot-$) and $K_q^{s}$ ($\cdots$) for the HSS are shown for comparison.  
The unit of length here and in the following figures is the hard-core
diameter $d$ of the particles.
}
\end{figure} 

The second prerequisite for the asymptotic description according to Eq.\
(\ref{eq:asy:G1G2q}) are the wave-vector dependent amplitudes $f_q^*$,
$h_q^*$ and $K_q^*$. These are shown for the $A_4$-singularity in Fig.\
\ref{fig:swsasy:fqA4} together with the related values for the
tagged-particle correlator, Eq.\ (\ref{eq:asy:phis}).  The quantities for
the tagged particle motion are close to the ones for the coherent
correlator $\phi_q(t)$ except for values of $q$ smaller than, say, $q=10$.
This difference was observed already for the HSS \cite{Fuchs1998}. Since
we will not be concerned with small $q$ in the following, we restrict the
discussion to the coherent dynamics and imply that the same is applicable
also to the incoherent part with only minor changes. In comparison to the
HSS the $f_q^*$, $h_q^*$, $f^{s\,*}_q$ and $h^{s\,*}_q$ are extended over
a broader $q$-range. The maximum in $h^s_q$ is shifted from $q\approx 13$
to $q\approx 25$ reflecting the smaller localization length in the SWS as
noticed before, cf. \cite{Dawson2001}. We see in the lower panel of Fig.\
\ref{fig:swsasy:fqA4} that the distributions of the correction amplitudes
$K_q$ and $K^s_q$ share that trend of becoming broader from the HSS to the
$A_4$-singularity of the SWS.  The zero in $K_q$ moves from around
$q\approx 14$ in the HSS to $q\approx 32$ in the SWS. In addition, the
amplitudes are shifted to lower values for small $q$.

\begin{figure}[htb] 
\includegraphics[width=\columnwidth]{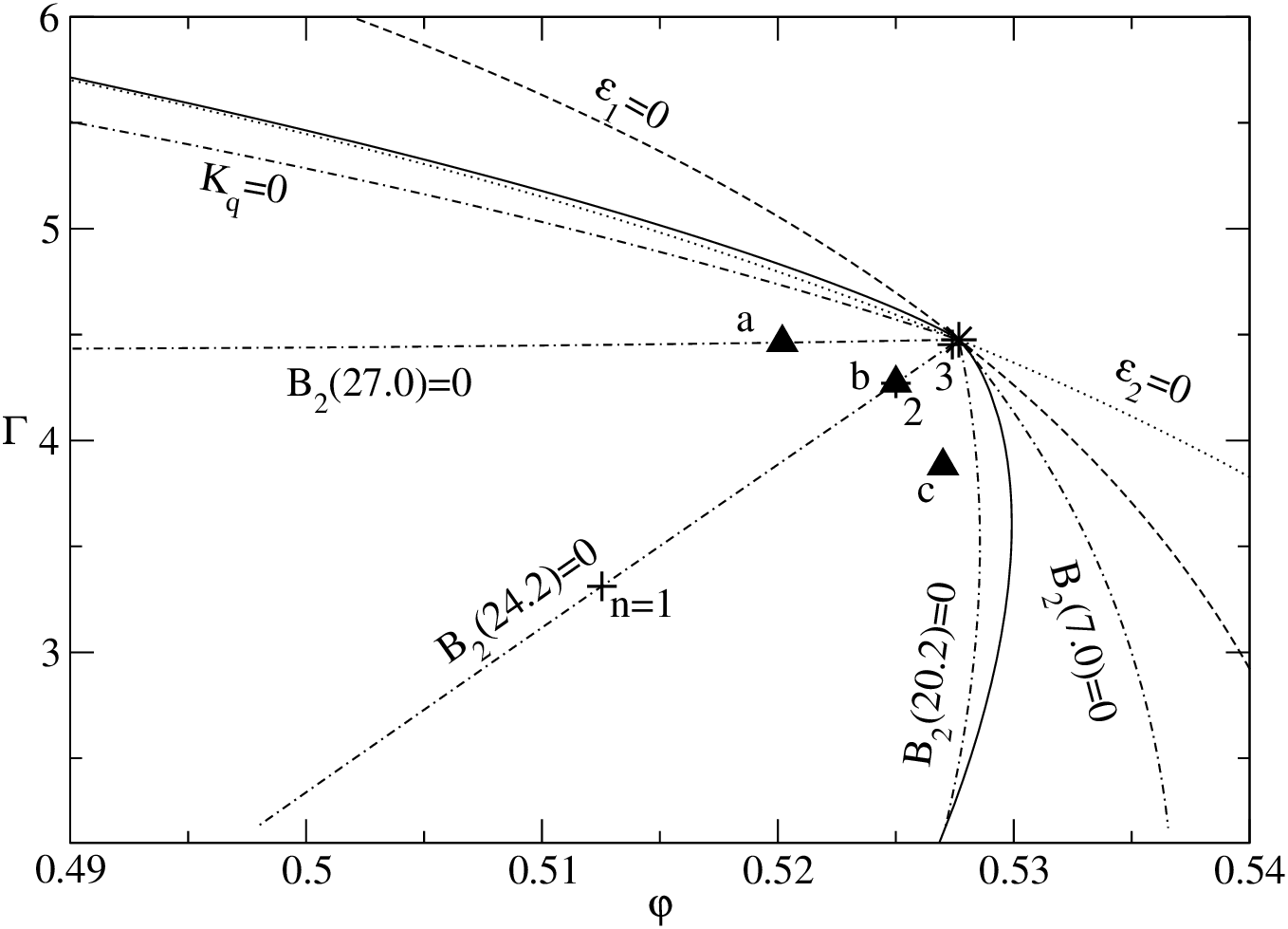}
\caption{\label{fig:swsasy:A4quadlines}Curves of vanishing quadratic 
correction in Eq.\ (\ref{eq:asy:G1G2q}) at the $A_4$-singularity of the 
SWS, $B_2(q)=0$ (dash-dotted), for $q=7.0$, $20.2$, $24.2$, $27.0$, and 
for $K_q=0$ as labeled. 
The full line shows a part of the glass-transition diagram for constant
$\delta=\delta^*$.  The lines of vanishing separation parameters
$\varepsilon_1(\mathbf{V})$ and $\varepsilon_2(\mathbf{V})$ are shown by a
broken and a dotted line, respectively. For the wave vector $q=24.2$, a
path on the curve $B_2(24.2)=0$ is marked ($+$) and labeled by $n$, for
which the correlators are shown in Fig.\ \ref{fig:swsasy:A4log}. State
$n=2$ is analyzed also in Fig.\ \ref{fig:swsasy:A4logqvar}. For the points
($\blacktriangle$) labeled a, b, and c the decay is shown in Figs.\
\ref{fig:swsasy:A4logVvar} and \ref{fig:conclog}.
} 
\end{figure} 

Having specified the characteristic parameters for the $A_4$-singularity,
the solution at any point in the control-parameter space can be compared
to the asymptotic approximation as the control parameters are translated
into separation parameters $\varepsilon_1$ and $\varepsilon_2$. As done
for the schematic models in Ref.~\cite{Goetze2002}, we start by
determining the surfaces where the quadratic corrections in Eq.\
(\ref{eq:asy:G1G2q}) are zero, $B_2(q)=B_2+K_qB^2=0$. On these surfaces in
the control-parameter space the logarithmic decay is expected to show up
as straight line around the plateau $f_q^*$, as the cubic and quartic terms
in Eq.\ (\ref{eq:asy:G1G2q}) vanish because of $B_3=B_4=\mu_3=0$ at the
$A_4$-singularity, cf. Eq.\ (\ref{eq:asy:Bcoeffs_B3}). 
We get a different surface for each wave vector $q$ and show typical examples 
in Fig.\ \ref{fig:swsasy:A4quadlines} for a cut through the glass-transition
diagram for $\delta=\delta^*$. For $q=7.0$ one gets $K_q=-1.81$. The
solution of $B_2(\mathbf{V})=1.81\,B(\mathbf{V})^2$ yields the chain line
labeled $B_2(7.0)=0$ in Fig.\ \ref{fig:swsasy:A4quadlines} and is lying in
the arrested region close to the line of liquid-glass transitions. Since
the $K_q$ depend smoothly on $q$, the evolution of the curve where
$B_2(q)=0$, can be understood by inspecting the parameters $B$ and $B_2$.
The square $B^2$ is always positive and proportional to
$\varepsilon_1(\mathbf{V})$, cf. Eq.\ (\ref{eq:asy:log_decay}), therefore
$K_q\,B^2$ is proportional to $K_q\,|\varepsilon_1(\mathbf{V})|$ and
shares the sign of $K_q$. Inserting $\mu_3=0$ and $\zeta=0.1216$ into Eq.\
(\ref{eq:asy:Bcoeffs_B2}) yields
$B_2(\mathbf{V})=0.111\,\varepsilon_1(\mathbf{V})  
-0.152\,\varepsilon_2(\mathbf{V})$, which has to be positive to comply
with $B_2(q)=0$. The second separation parameter is negative,
$\varepsilon_2(\mathbf{V})<0$, below the dotted curve for
$\varepsilon_2=0$ in Fig.\ \ref{fig:swsasy:A4quadlines}. In addition, the
value $|\varepsilon_2(\mathbf{V})|$ on the line $B_2(7.0)=0$ is larger
than $|\varepsilon_1(\mathbf{V})|$ which we can also infer from the fact
that the line $\varepsilon_1=0$ is closer than the line $\varepsilon_2=0$.
We now chose a point on the line $B_2(7.0)=0$, keep the first separation
parameter fixed, say $\varepsilon_1=\varepsilon_1'$, and move to higher
values for $K_q$, e.g., for $q=20.2$ where $K_q=-0.966$. $B^2$ stays the
same and the term $K_qB^2$ increases. To ensure that $B_2(20.2)=0$, the
value $B_2(\mathbf{V})$ has to decrease. We can achieve that by moving
closer to the line $\varepsilon_2=0$. For fixed $\varepsilon_1'$ this
implies a shift to lower $\varphi$ and higher $\Gamma$. Consequently the
entire line, where $B_2(q)=0$, is rotating clockwise around the
$A_4$-singularity as $K_q$ increases. This is seen for the chain line
$B_2(20.2)=0$ in Fig.\ \ref{fig:swsasy:A4quadlines}. Since $K^s_q$ is
monotonic increasing with $q$ and $K_q$ has the same trend when neglecting
the small oscillations, Fig.\ \ref{fig:swsasy:fqA4}, the line
$B^s_2(q)=B_2(\mathbf{V}) +B(\mathbf{V})^2K^s_q=0$ also rotates clockwise
with increasing wave-vector $q$.

The variation of the lines $B_2(q)=0$ described above depends only on the
angle at which the lines $\varepsilon_1=0$ and $\varepsilon_2=0$ intersect
at the $A_4$-singularity. This intersection is in a sense generic that it
is shared by the close-by $A_3$-singularities of the SWS. It applies also
to the $A_4$-singularities of the other potentials which are similar to
the square well. This is so because the functionals determining the
separation parameters depend on quantities like the structure factors and
the glass-form factors which are similar for different potentials
\cite{Goetze2003b}. For a given wave vector $q$, the line $B_2(q)=0$ may
or may not lie in the liquid regime depending on $K_q$. For the SWS at
$\delta=\delta^*$ we get a range of $-1\lesssim K_q\lesssim 0.4$
corresponding to $20\lesssim q \lesssim 35$, where a line $B_2(q)=0$ is
found in the liquid regime. We illustrate this by adding lines for
$q=24.2$, $q=27.0$ and for $K_q=0$ to Fig.\ \ref{fig:swsasy:A4quadlines}.
The vanishing $K_q$ is corresponding to $q\approx 32.3$ yielding a line 
$B_2(q)=0$ still in the liquid. For $q\gtrsim 35$, the latter line rotates
further around the $A_4$-singularity and into the arrested regime beyond the 
almost horizontal line of liquid-glass transitions.

\begin{figure}[htb] 
\includegraphics[width=\columnwidth]{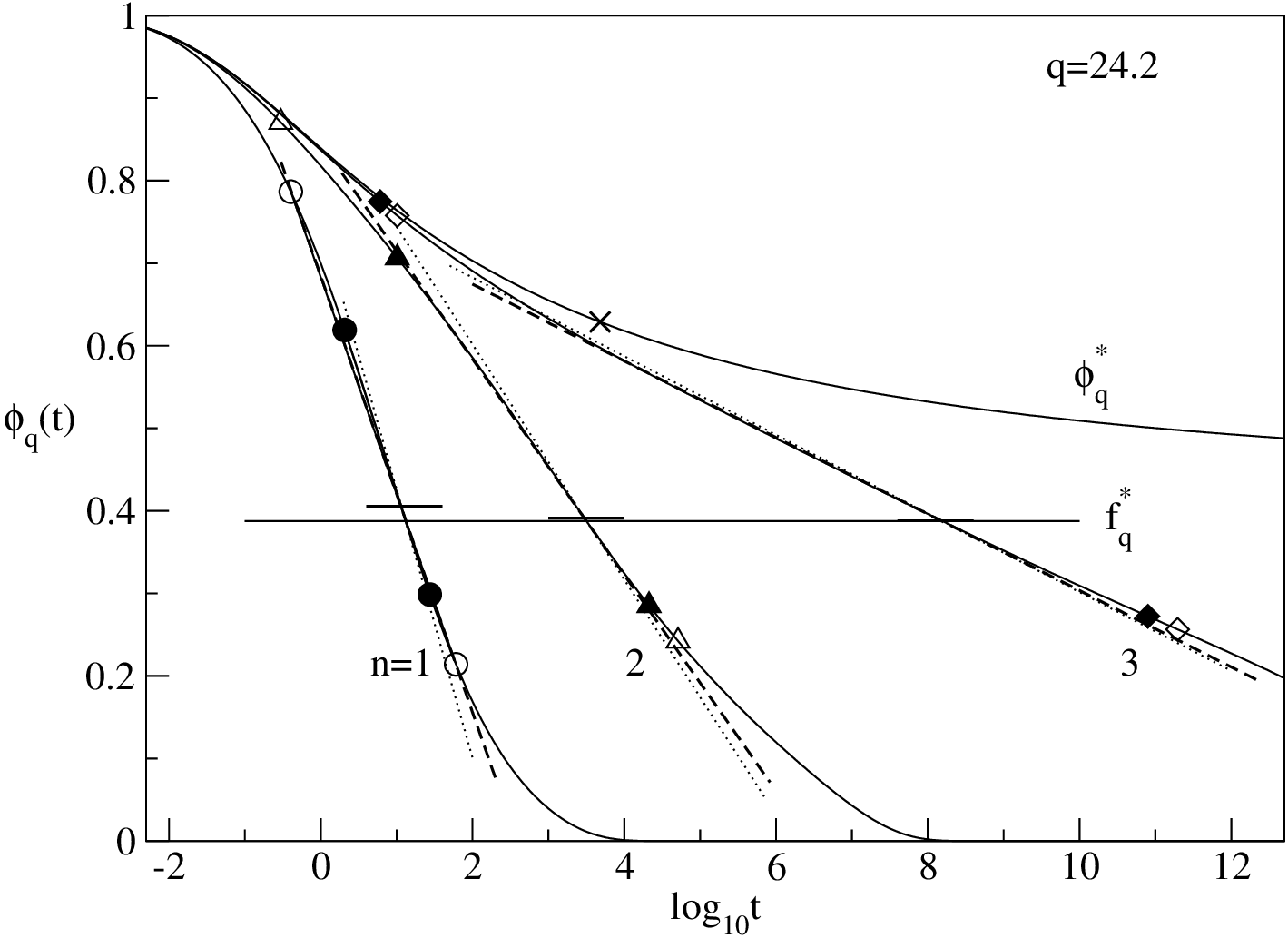}
\caption{\label{fig:swsasy:A4log}Logarithmic decay at the $A_4$-singularity 
in the SWS for $q=24.2$ on the path indicated in 
Fig.\ \ref{fig:swsasy:A4quadlines}.
The correlation functions are shown as full lines for the states
$n=1,\,2,\,3$ (see text) and at $\mathbf{V}=\mathbf{V}^*$ labeled
$\phi_q^*$. The horizontal line indicates the critical plateau value
$f_q^*$ for $q=24.2$, short lines the renormalized plateaus
$f_q^*+\hat{f}_q$. Broken lines show the approximation of Eq.\
(\ref{eq:asy:G1G2q}), $-(B-B_1)\,\ln(t/\tau)$, dotted lines the
approximation by Eq.\ (\ref{eq:asy:log_decay}). Filled and open symbols,
respectively, mark the points where the approximations deviate by $5\%$
from the solution. The cross indicates the time when the solution for
$n=3$ and the critical correlator $\phi_q^*$ differ by $5\%$. The unit of
time here and in the following figures is given by a short-time diffusion
coefficient of $D_0=1/160$.
} 
\end{figure}

We select a wave vector $q=24.2$ with $K_q=-0.596$ as indicated in Fig.\
\ref{fig:swsasy:fqA4} by a square and choose a path on the line
$B_2(24.2)=0$ marked by the plus symbols in Fig.\
\ref{fig:swsasy:A4quadlines}. For $n=1,\,2,\,3$, the control parameters
are $(\Gamma,\varphi)=(3.312,0.5125)$, $(4.271,0.5250)$, and
$(4.453,0.5274)$, respectively. The solutions are shown in Fig.\
\ref{fig:swsasy:A4log} together with the leading approximation, Eq.\
(\ref{eq:asy:log_decay}), (dotted) and the next-to-leading approximation,
Eq.\ (\ref{eq:asy:G1G2q}), (dashed). The time scales $\tau$ are matched at
the plateau $f_q^*$ for the leading approximation and at the renormalized
plateaus $f_q^*+\hat{f}_q$ for the first correction. We recognize that for
$n=3$, Eq.\ (\ref{eq:asy:G1G2q}) accounts for more than ten decades in
time with a relative accuracy better than $5\%$. The leading approximation is
acceptable on that level for nine decades. For $n=1$ two and more than one
decade are covered, respectively. Five and three orders of magnitude in
time are achieved for $n=2$. For $n=1,\,2,\,3$, the leading approximation
describes at least $30\%$ of the complete decay and when including the
correction, $65\%$ are covered on the chosen accuracy level. The distance
in the control parameter $\Gamma$ from the value at the $A_4$-singularity
is $25\%$ for $n=1$ and $4\%$ for $n=2$, so no fine-tuning was necessary
to obtain such large windows for the logarithmic decay. The curve $n=1$
requires about five decades for the complete decay which is well in the
reach of today's computer simulation techniques \cite{Kob2003pre}.

It was possible to describe part of the critical decay at an
$A_3$-singularity in a one-component model by the expansion in polynomials
in $\ln t$ at a point away from the singularity \cite{Goetze2002}. We
therefore compare the critical decay $\phi_q^*(t)$ with the decay for
$n=3$ and indicate the point at $t\approx 5000$ where both differ by $5\%$
in Fig.\ \ref{fig:swsasy:A4log}. With only the leading correction at our
disposal, a $2\%$-criterion was not fulfilled as for the one-component
model, where also the next-to-leading correction could be used
\cite{Goetze2002}. The dashed line for $n=3$ does not come closer to the
critical decay than $4\%$. Allowing for $5\%$, the interval from $t\approx
20$ to $t\approx 4000$ could be described. However, at the
$A_4$-singularity the approximation in Eq.\ (\ref{eq:asy:G1G2q}) always
yields a straight $\ln t$-decay as approximation on the chosen path with
$B_2(q)=0$.  This disagrees qualitatively with the observed critical
decay.

\begin{figure}[htb] 
\includegraphics[width=\columnwidth]{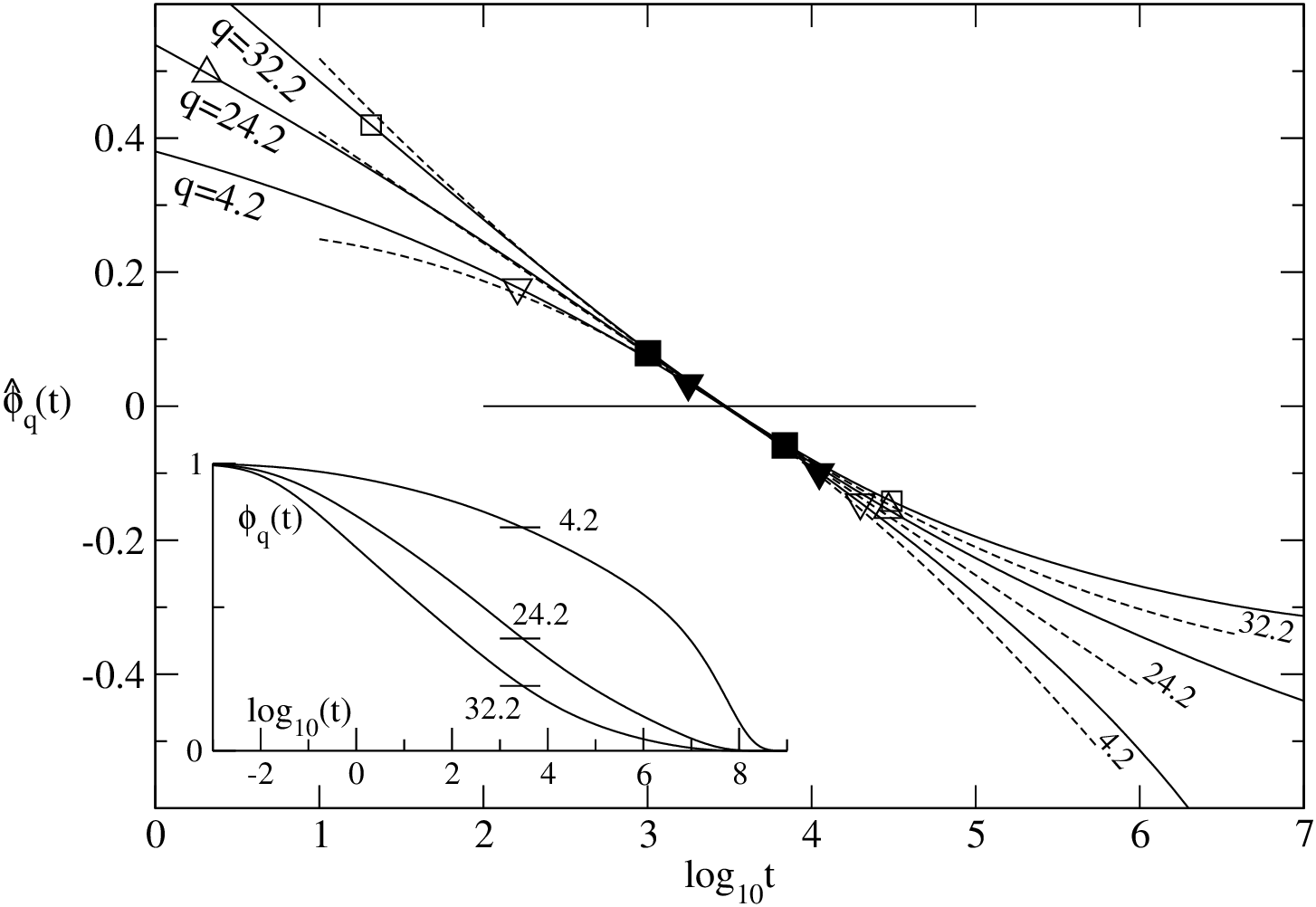}
\caption{\label{fig:swsasy:A4logqvar}Logarithmic decay at the 
$A_4$-singularity for varying wave vector $q$. 
The inset shows the correlation functions $\phi_q(t)$ at state $n=2$ from
Fig.\ \ref{fig:swsasy:A4quadlines} for wave vectors $q=4.2,\,24.2\,,$ and
$32.2$ from top to bottom and the short horizontal lines show the
corresponding critical plateau values $f_q^*$. The full panel shows the
same correlation functions rescaled according to $\hat{\phi}_q(t)=
(\phi_q(t)-f^*_q-\hat{f}_q)/h^*_q$ as full lines and labeled by the
respective wave vectors. Dashed lines show the asymptotic laws, Eq.\
(\ref{eq:asy:G1G2q}). The deviations of the approximations from the
solutions of $5\%$ are marked by the open symbols. Filled symbols for
$q=4.2$ ($\blacktriangledown$) and $q=32.2$ ($\blacksquare$) show the
$5\%$ deviation from the additional approximation of neglecting quadratic
terms in Eq.\ (\ref{eq:asy:G1G2q}) (see text).
} 
\end{figure} 

To identify correctly some decay that is linear in the $\phi_q(t)$ versus
$\log t$ diagram with the logarithmic decay predicted by the asymptotic
laws, Eq.\ (\ref{eq:asy:G1G2q}), we check if a different correlator with a
different correction amplitude $K_q$ is \textit{not} linear in $\ln t$ at
the same point in the control-parameter space. For a two-component model a
characteristic alternation of concave, linear and convex decay in $\ln t$
was found \cite{Goetze2002,Sperl2003}. Not both correlators could be
linear in $\ln t$ at the same time. For the SWS this check is performed at
the point $n=2$ from Fig.\ \ref{fig:swsasy:A4quadlines} by variation of
the wave vector $q$. For the wave vectors $q=4.2$ and $32.2$ the
correction amplitudes are $K_q=-1.400$ and $-0.0413$, respectively.
Therefore $B_2(4.2)<0$ and $B_2(32.2)>0$. We expect $\phi_q(t)$ to be
concave or convex, accordingly, as is demonstrated by the inset of Fig.\
\ref{fig:swsasy:A4logqvar}. The rescaled correlators $\hat{\phi}_q(t)$
displayed in the full panel allow for a more detailed analysis. We see
that the solutions as well as the approximations clearly exhibit increased
curvature for larger $q$. Since the coefficient linear in $\ln t$ is not
depending on $q$, cf. Eq.\ (\ref{eq:asy:G1G2q}), the middle dashed line
represents the leading correction to all three correlators when the
quadratic terms are neglected. For $q=24.2$ we observe good agreement over
almost 5 decades as before, cf. Fig.\ \ref{fig:swsasy:A4log}. For $q=4.2$
and $32.2$, however, the additional approximation reduces the range of
applicability to less than one decade as marked by the filled symbols.
Including the quadratic terms from the approximation~(\ref{eq:asy:G1G2q})
extends this range by half a decade to later times and to earlier times by
one and almost two decades for $q=4.2$ and $32.2$, respectively. The time
window defined by a $5\%$ deviation from the
approximation~(\ref{eq:asy:G1G2q}) is larger by one and two orders of
magnitude for $q=24.2$ than for $q=32.2$ and $q=4.2$, respectively, what
indicates that $q$-dependent higher-order corrections significantly
influence the range of applicability for the leading
correction~(\ref{eq:asy:G1G2q}).

The time scale $\tau$ in Fig.\ \ref{fig:swsasy:A4logqvar} was matched for
$q=24.2$, so the violation of scale universality inherent to an
approximation like in Eq.\ (\ref{eq:asy:G1G2q}) leads to different times
$\tau(q)$, where the correlators for different $q$ cross their respective
renormalized plateau $f_q^*+\hat{f}_q$ \cite{Goetze2002}. The
representation with the rescaled $\hat{\phi}_q(t)$ is particularly
sensitive to these deviations since the point where the plateau is crossed
is required to be zero, $\hat{\phi}_q(t/\tau)=0$. In Fig.\
\ref{fig:swsasy:A4logqvar} we see that the line crossing the zero is
slightly broader than a single curve. The deviations in $\tau(q)$ are
small enough to not exceed the numerical grid for the time axis which
around $\tau=2988$ is given by $\Delta t=172$.  So we interpolate to get
for $q=4.2,\,24.2$, and $32.2$, $\tau(q)=2899$, $2988$, and $3017$,
respectively. These differences do not introduce larger errors in the
analysis carried out above.

\begin{figure}[htb]
\includegraphics[width=\columnwidth]{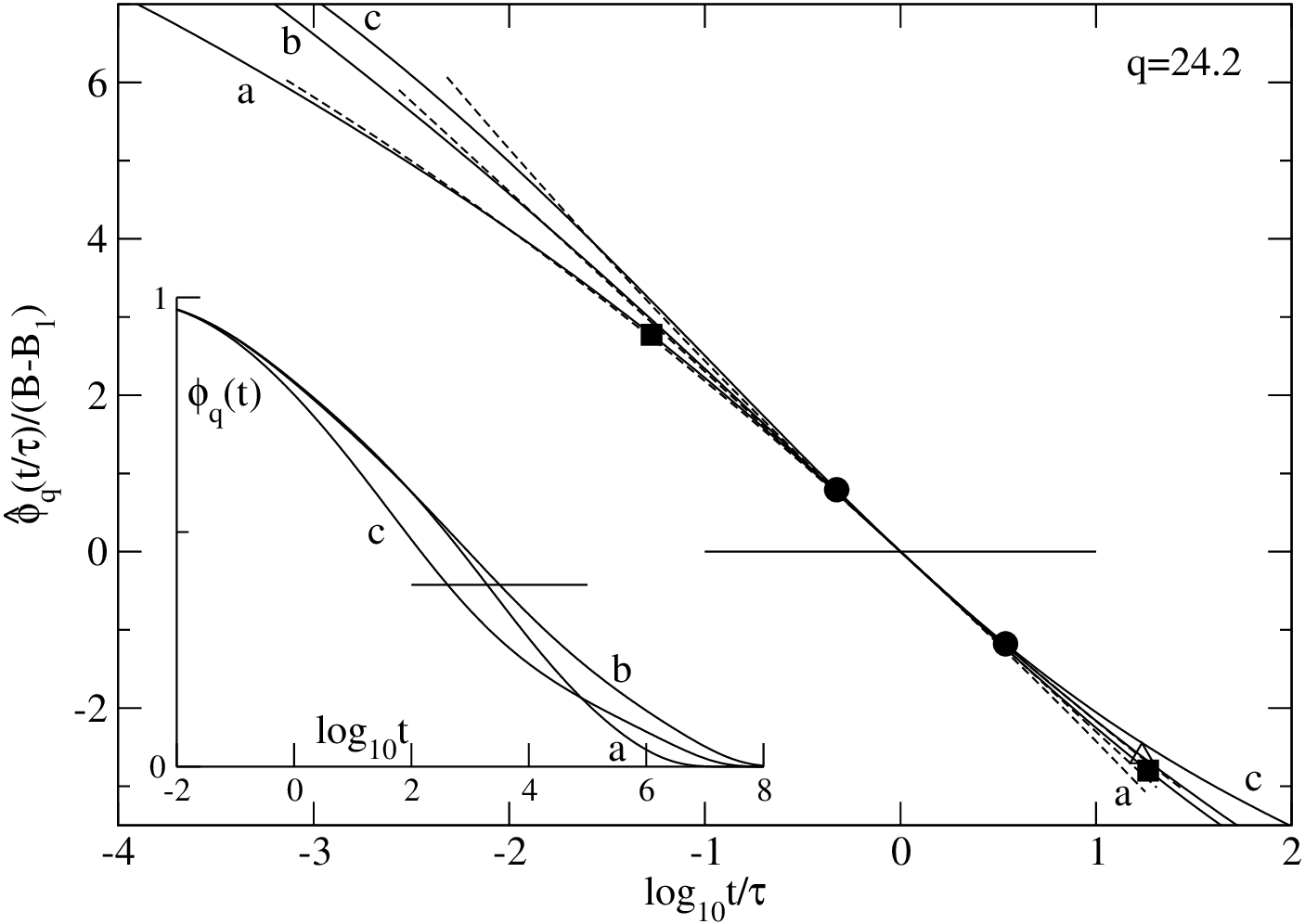}
\caption{\label{fig:swsasy:A4logVvar}Logarithmic decay at the 
$A_4$-singularity for the three states marked by triangles in 
Fig.\ \ref{fig:swsasy:A4quadlines}. 
The inset shows the correlation functions $\phi_q(t)$ for $q=24.2$. The
plateau value $f_q^*$ is indicated by the short horizontal line. The full
panel shows $\hat{\phi}_q(t)=(\phi_q(t)-f_q^*-\hat{f}_q)/h_q$ divided by
the respective values for $(B-B_1)$ at the three states specified. The
dashed curves show the result from Eq.\ (\ref{eq:asy:G1G2q}). Filled
squares and circles mark the points where curve a and c deviate by $5\%$
from $-\ln(t/\tau)$, respectively.  The deviation for curve~b
($\triangle$) for short times is at $t/\tau\approx 10^{-4}$ and not
included in the figure.
}
\end{figure}

In order to change from convex to concave behavior we can also change the
control parameters. For states above the line $B_2(24.2)=0$, we expect
concave behavior, $B_2(q)<0$, for states below, convex decay,
$B_2(24.2)>0$. For a demonstration of this result, the rescaled
correlators $\hat{\phi}_q(t/\tau)$ at the states labeled a, b, c in Fig.\
\ref{fig:swsasy:A4quadlines} are divided by the prefactor $(B-B_1)$ of
Eq.\ (\ref{eq:asy:G1G2q}). This way the part of the decay that is linear
in $\ln t$ shows up as straight line with slope $-\ln 10$ in Fig.\
\ref{fig:swsasy:A4logVvar}. The approximations~(\ref{eq:asy:G1G2q}) are
shown as dashed lines for each state representing
$-\ln(t/\tau)+[B_2(24.2)/(B-B_1)]\ln^2(t/\tau)$. For state~b the
approximation is identical to $-\ln(t/\tau)$ and the solution follows that
line over 5 decades before $5\%$ deviation is reached. The states a and c
are chosen to have the same value for $B-B_1\approx 0.015$ and
$B_2(24.2)=\mp0.0020$, respectively. The solutions at state~a and~c follow
the $-\ln t$-law closely within a $5\%$ margin for two decades or one
decade, respectively, which is significantly less than found for state~b.
We can infer from Fig.\ \ref{fig:swsasy:A4quadlines} that at state~a the
quadratic corrections would vanish again if we went from $q=24.2$ to the
higher wave vector $q=27.0$. A scenario similar to the one shown in Fig.\
\ref{fig:swsasy:A4log} can be found.

The procedure outlined in Figs.~\ref{fig:swsasy:A4quadlines},
\ref{fig:swsasy:A4log}, \ref{fig:swsasy:A4logqvar}, and
\ref{fig:swsasy:A4logVvar} can be summarized as follows. From the
higher-order singularities there emanate surfaces in the control-parameter
space for a specific wave vector $\bar{q}$ where the quadratic term in
Eq.\ (\ref{eq:asy:G1G2q}) is zero, cf. Fig.\ \ref{fig:swsasy:A4quadlines},
and the decay is linear in $\ln t$. Moving closer to the singularity on
that surface, the window in time where the logarithmic decay is a valid
approximation increases, cf. Fig.\ \ref{fig:swsasy:A4log}. On a fixed
point on that surface the decay is concave for $q<\bar{q}$ and convex for
$q>\bar{q}$, cf. Fig.\ \ref{fig:swsasy:A4logqvar}. For fixed $\bar{q}$,
the change from concave to convex is achieved by crossing the mentioned
surface from above in the sense exemplified in Fig.\
\ref{fig:swsasy:A4logVvar}.

The coupled quantities share the leading asymptotic behavior of the
density correlators. As a consequence of the factorization theorem of MCT,
only the glass-form factors and the critical amplitudes $h_q$ are
different for the coupled quantities \cite{Goetze1985}. The leading
corrections imply a violation of a generalized factorization theorem.
These are proportional to the correction amplitude $K_q$. Since for large
wave vectors, say $q>10$, the quantities $f^s_q$, $h^s_q$, and $K^s_q$ are
close to the ones for the coherent correlator, the approximation for the
tagged particle correlation functions $\phi^s_q(t)$ for these large $q$ is
the same as for $\phi_q(t)$. So the discussion for $\phi^s_q(t)$ is
already exhausted by Fig.\ \ref{fig:swsasy:fqA4}. Not much could be gained
from repeating the discussion of the previous section for $\phi^s_q(t)$.

\section{\label{sec:MSD}Mean Squared Displacement near an $A_4$-singularity}

\begin{figure}[htb]
\includegraphics[width=\columnwidth]{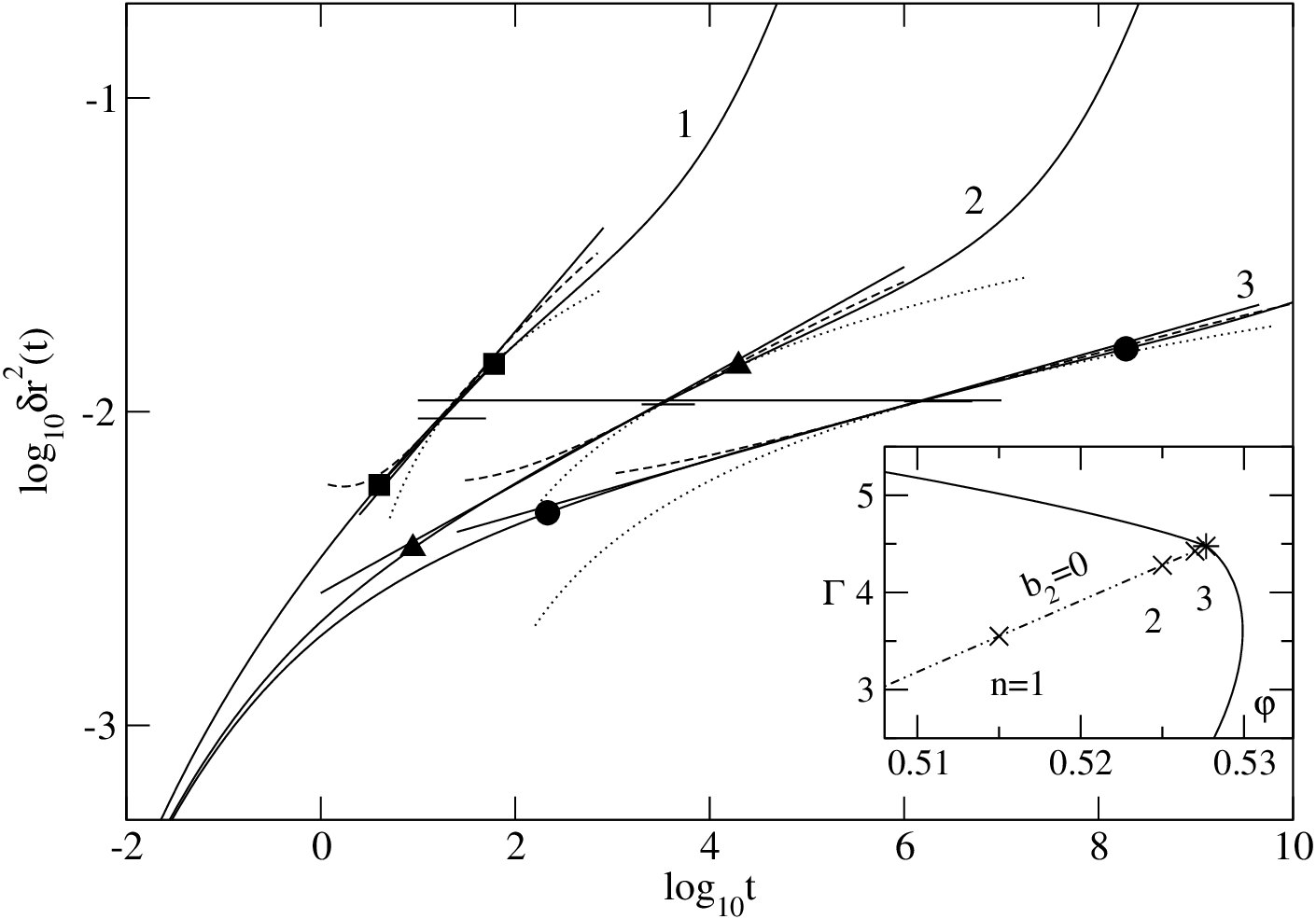}
\caption{\label{fig:swsasy:MSDx}Subdiffusive power law in the mean-squared 
displacement (MSD). 
The solutions for states 1, 2, and 3 in the inset are shown as full lines
in the full panel together with the leading (dotted) and next-to-leading
(dashed) approximation by Eq.~(\ref{eq:asy:MSD}). The long horizontal line
represents $6\,r_s^{*\,2}=0.01086$, the short horizontal lines the
corrections to the plateau, $6(r_s^{*\,2}-\hat{r}_s^2)$, cf. Eq.\
(\ref{eq:asy:MSDpar:df}). The straight full lines show the power law
$(t/\tau)^{x}$, Eq.\ (\ref{eq:asy:expoxlaw}), with exponents $x=0.365$,
$0.173$ and $0.0878$ for states $n=1,\,2,\,3$. The filled symbols show the
points where the solutions deviate by $5\%$ from the leading-order power
laws. The inset shows part of the glass-transition diagram for
$\delta=\delta^*$ and a chain line where $b_2=0$, 
cf. Eq.\ (\ref{eq:asy:expoxpar}), (see text).
}
\end{figure}

According to Eq.\ (\ref{eq:asy:expoxexp}), $\delta r^2(t)$ is expected to
exhibit power-law behavior around the plateau $6\,r_s^{c\,2}$ provided the
term $b_2$ vanishes. The power-law exponent $x$ is determined explicitly
in Eq.\ (\ref{eq:asy:expoxcalc}) by the localization length and the
critical amplitude, which are
\begin{equation}\label{eq:swsasy:A4rs}
r_s^{*}=0.04255\,,\quad h^*_\text{MSD}=0.004051 \,.
\end{equation}

The inset of Fig.\ \ref{fig:swsasy:MSDx} shows the line where $b_2$ from
Eq.\ (\ref{eq:asy:expoxexp}) vanishes. This line is almost identical to
the one for $B_2(24.2)=0$ shown in Fig.\ \ref{fig:swsasy:A4quadlines} for
the correlators $\phi_q(t)$. The MSD for three states on that line is
shown in the full panel. It is described well by the approximation in Eq.\
(\ref{eq:asy:MSD}). For states $n=1,\,2,\,3$, one, three and six decades
are covered with deviations less than $5\%$, so the approximation yields a
description of similar accuracy as for the correlation functions in Fig.\
\ref{fig:swsasy:A4log}. The leading result from Eq.\
(\ref{eq:asy:log_decay}) describes the relaxation proportional to $\ln t$
(dotted) which always has negative curvature in the double-logarithmic
representation and does not provide a valid description for $n=1$ and 2.
The reason for the qualitative difference between the solution for the MSD
and the leading logarithmic law is that the corrections proportional to
$K_\text{MSD}=-1.708$ are large, Therefore, $K_\text{MSD}B^2+B_2$ is never
close to zero in the liquid regime except very close to the
$A_4$-singularity. This is seen for $n=3$ in Fig.\ \ref{fig:swsasy:MSDx}
where $\ln t$ develops a straightened decay around the plateau.

The power law~(\ref{eq:asy:powerlead}) provides a different formulation of
a leading order approximation and is shown in Fig.\ \ref{fig:swsasy:MSDx}
as straight line for $n=1,\,2,\,3$.  For $n=1$ this describes the MSD for
more than a decade as indicated by the squares. For $n=2$ three decades
are covered and six decades of power-law behavior are identified for curve
$n=3$. So the accuracy is similar to the one provided by the approximation
in next-to-leading order by Eq.\ (\ref{eq:asy:MSD}). Both asymptotic
descriptions fall on top of each other around the plateau and therefore
corroborate that the reformulation~(\ref{eq:asy:expoxexp}) is justified.
The interpretation of the behavior of the MSD is then much simpler when
considering the power laws instead of the logarithms of time.  The
decreasing slope of the relaxation when approaching the $A_4$-singularity
as in Fig.\ \ref{fig:swsasy:MSDx} is just the exponent $x$ from Eq.\
(\ref{eq:asy:expoxcalc}) which decreases as $B$ with the square-root of
the separation parameter $\varepsilon_1$, cf.  Eq.\
(\ref{eq:asy:log_decay}). The same parameter $B$ is the prefactor of the
leading-order logarithmic decay in Eq.\ (\ref{eq:asy:log_decay}). In that
sense Fig.\ \ref{fig:swsasy:MSDx} is the analog of Fig.\
\ref{fig:swsasy:A4log}.

\begin{figure}[htb]
\includegraphics[width=\columnwidth]{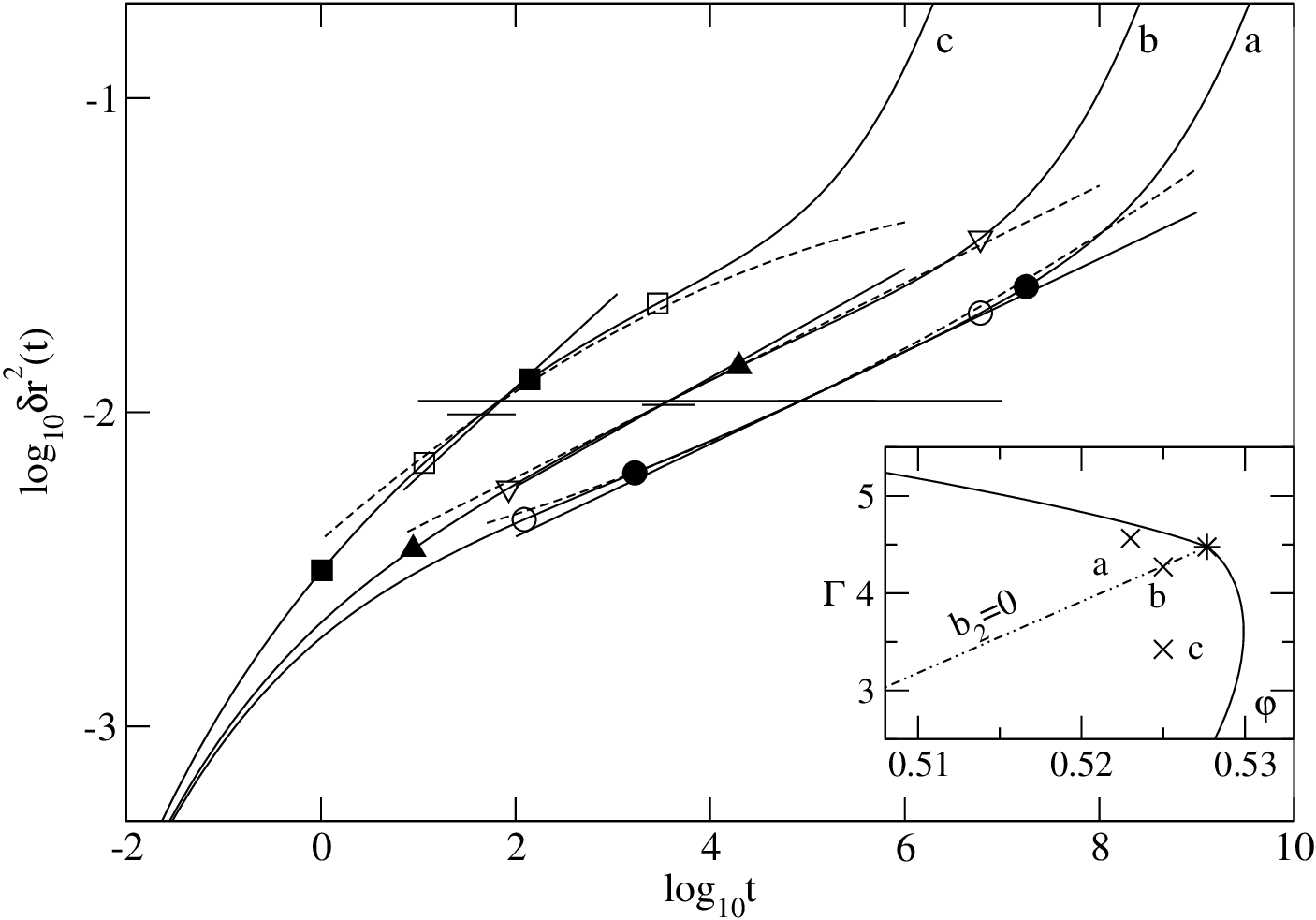}
\caption{\label{fig:swsasy:MSDV}Concave and convex deviations from the 
power law, Eq.\ (\ref{eq:asy:powerlead}) in the MSD. 
Solutions for the states a, b, and c are shown as full lines, the
approximation~(\ref{eq:asy:powerlead}) as straight full lines for
exponents $x=0.147$, $0.173$, and $0.285$, respectively. Filled symbols
denote the $5\%$ deviation of the solutions from the leading-order power
law. For state~b, the dashed line exhibits the corrected power law with
$x'=0.155$, Eq.\ (\ref{eq:asy:b1dash}), and the open triangle the $5\%$
deviations of the solution from it. Dashed lines show the approximation by
Eq.\ (\ref{eq:asy:powercorrb2}) for a and c with $b_2=0.00363$ and
$-0.00735$, respectively, and $x'=0.143$ and $0.214$. The open symbols
mark the $5\%$ deviations. The inset replots the one from Fig.\
\ref{fig:swsasy:MSDx} and shows by the crosses the state points a, b, and
c.
}
\end{figure}

The term $b_2$ in Eq.\ (\ref{eq:asy:expoxexp}) varies regularly in the
separation parameters $\varepsilon_1$ and $\varepsilon_2$, and $b_2$ is
positive above the line $b_2=0$ and negative below. Therefore, similar to
the case for the correlators in the linear-$\log$ representation, in the
double-logarithmic representation, the behavior of the MSD can be changed
from convex to concave when crossing the line of vanishing $b_2$. This is
demonstrated for three states in Fig.\ \ref{fig:swsasy:MSDV}. State~b is
identical to the state $n=2$ in Fig.\ \ref{fig:swsasy:MSDx} and obeys
$b_2=0$. The power law $(t/\tau)^{x}$ is shown as straight full line. The
time scale $\tau$ is matched at the plateau $6\,r_s^{*\,2}$. Moving to
state~c below the chain line, $(\Gamma, \varphi)=(3.42,0.525)$, a
relaxation is obtained which clearly exhibits negative curvature and is
consistent with the calculated value $b_2=-0.00735$. The leading-order
power law with exponent $x=0.285$ fulfills a $5\%$-deviation criterion for
two decades which accidentally extends to short times as the approximation
crosses the solution twice. Reducing the allowed deviation to $4\%$ would
reduce that interval to less than a decade. If we include the term
proportional to $b_2$ from Eq.\ (\ref{eq:asy:powercorrb2}) and renormalize
the exponent to $x'$, Eq.\ (\ref{eq:asy:b1dash}), the approximation agrees
with the solution for three decades. It is obvious from a comparison with
curve~1 in Fig.\ \ref{fig:swsasy:MSDx}, that the leading-order power law
describes that solution better than it describes the solution at state~c
in Fig.\ \ref{fig:swsasy:MSDV} for comparable values for $\tau$ and the
plateau correction $\hat{r}^2_{s}$ . The deviation to convex behavior is
demonstrated by the dashed line at curve~a,
$(\Gamma,\varphi)=(4.57,0.523)$. Again the range of validity is extended
to earlier times but for later times no improvement can be found.

In Fig.\ \ref{fig:swsasy:MSDx} the dashed line, which describes the
next-to-leading order approximation of Eq.\ (\ref{eq:asy:G1G2q}), deviates
from the leading order power law~(\ref{eq:asy:expoxlaw}) below the plateau
where the range of validity for the power law extends to much smaller
times than justified by its derivation.  We also recognize that the
exponent $x$ overestimates the slope of the relaxation in
Figs.~\ref{fig:swsasy:MSDx} and~\ref{fig:swsasy:MSDV}.  In Eq.\
(\ref{eq:asy:expoxcalc}) only the term $B$ from the leading order
approximation is present. Taking into account the renormalization of this
prefactor to $B-B_1$ in Eq.\ (\ref{eq:asy:b1dash}) changes the exponent
for state b from $x=0.173$ to $x'=0.155$. By comparing the full line for
the leading result and the dashed line for the corrected one in Fig.\
\ref{fig:swsasy:MSDV}, we find that the range of applicability is shifted
to later times by one decade and extended by two decades. The corrected
power law is valid from $t=10^2$ to $t=5\cdot10^6$ and comparison to Fig.\
\ref{fig:swsasy:MSDx} shows that approximation~(\ref{eq:asy:G1G2q}) covers
a similar range. The accidental extension to shorter times is removed. The
approximation now covers the range also a naive power-law fit would yield.

In summary, the correction amplitude $K_\text{MSD}$ for the MSD does not
vanish within the liquid regime. Therefore a logarithmic relaxation law
can be detected only for states very close to the singularity. However,
there is a line of vanishing corrections for the logarithm of the MSD.
Here a logarithmic relaxation can be observed and this describes a
subdiffusive power law of the MSD. We can interpret Fig.\
\ref{fig:swsasy:MSDV} as the analog of Fig.\ \ref{fig:swsasy:A4logVvar}.
Some quadratic correction to a leading-order linear behavior can be set to
zero on a surface in control-parameter space. Departing from that surface
in opposite directions introduces either positive or negative corrections
and the linear behavior is changed to convex or concave.

\section{\label{sec:A3}$A_3$-singularities}

\begin{figure}[htb]
\includegraphics[width=\columnwidth]{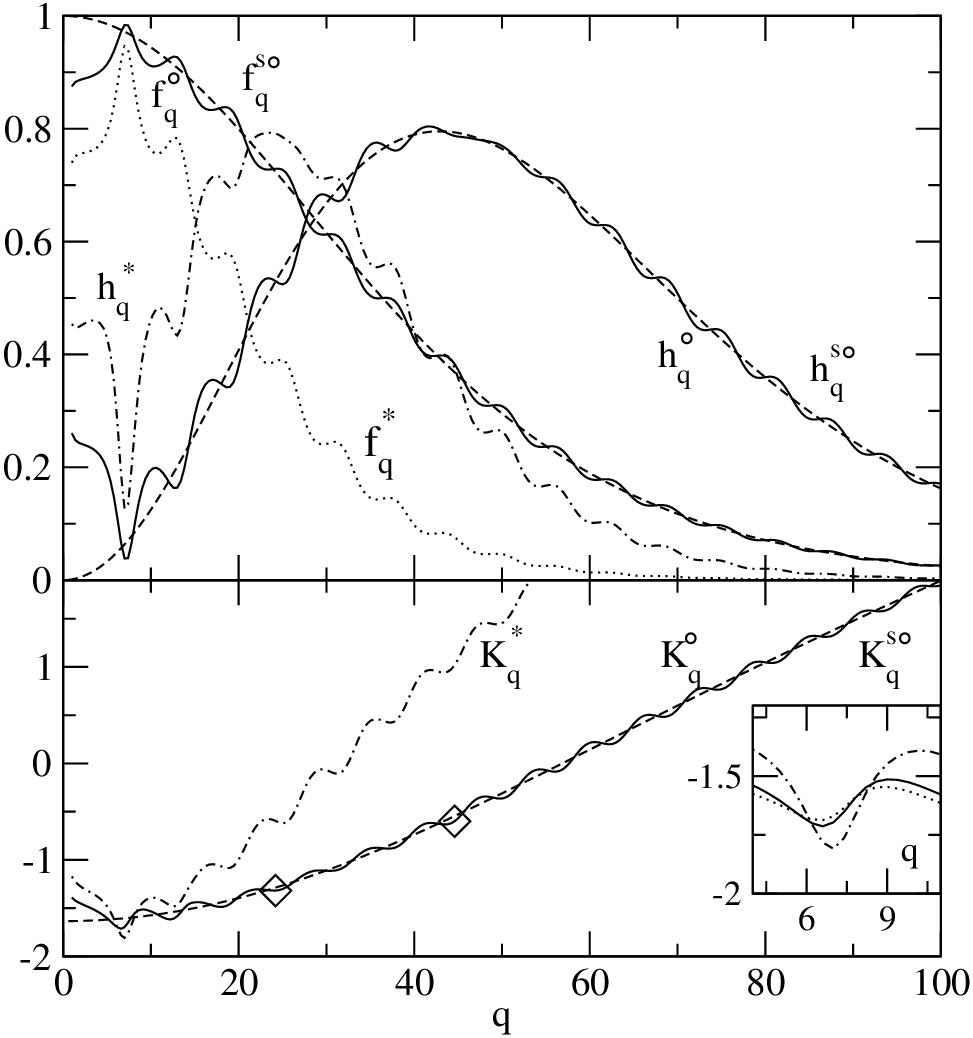}
\caption{\label{fig:swsasy:fqA3}Glass-form factors $f_q^\circ$ and 
$f_q^{s\,\circ}$, amplitudes $h_q^\circ$ and $h_q^{s\,\circ}$, and 
correction amplitudes $K_q^\circ$ and $K_q^{s\,\circ}$ at the 
$A_3$-singularity for $\delta=0.03$. 
Line styles are the same as in Fig.\ \ref{fig:swsasy:fqA4}. The values at
the $A_4$-singularity, $f_q^*$ (dotted), $h_q^*$ (dash-dotted), and
$K_q^*$ (dash-dotted), are shown for comparison.  The values for $q=24.2$
and $45.0$ are marked by diamonds. The inset shows $K_q$ for $4<q<11$ for
$\delta=\delta^*$ (chain line), $0.03$ (full line) and $0.02$ (dotted
line).
}
\end{figure}

An $A_3$-singularity is not located on a liquid-glass-transition line but
is the endpoint of a glass-glass-transition line \cite{Goetze1991b}. The
parameter $\mu_3$ is no longer vanishing and for $\delta=0.03$ we get
$\mu_3=0.109$ and $\zeta=0.157$. For this $A_3$-singularity the
$q$-dependent amplitudes are shown in Fig.\ \ref{fig:swsasy:fqA3}. No
qualitative changes are obvious compared to the results shown in Fig.\
\ref{fig:swsasy:fqA4} for the $A_4$. The smaller length scale
$\delta=0.03$ for the attractive well introduces a smaller localization
length, and this implies the broader distributions in wave-vector space.
So the trend seen when changing from the HSS to the $A_4$-singularity of
the SWS is continued when approaching $A_3$-singularities at smaller
$\delta$. There are only two notable exceptions at smaller $q$. First, the
value for $K_q$ at the position of the structure factor peak is minimal
for the $A_4$, $-1.81=K_q^*<K_q^\circ=-1.72$. The inset shows this region
enlarged for $\delta=\delta^*$, $0.03$ and $0.02$, demonstrating that
$K_q$ at the peak is again larger for the $A_3$-singularity with smaller
well width $0.02$, where $K_q=-1.69$. Second, the zero-wave-vector limit
of $K^s_q$ is also smallest at the $A_4$-singularity. The respective
values for $\delta=\delta^*$, $0.03$ and $0.02$ are $-1.71$, $-1.64$, and
$-1.62$. Therefore, one experiences the strongest $q$-dependent
corrections at the $A_4$-singularity.

\begin{figure}[htb]
\includegraphics[width=\columnwidth]{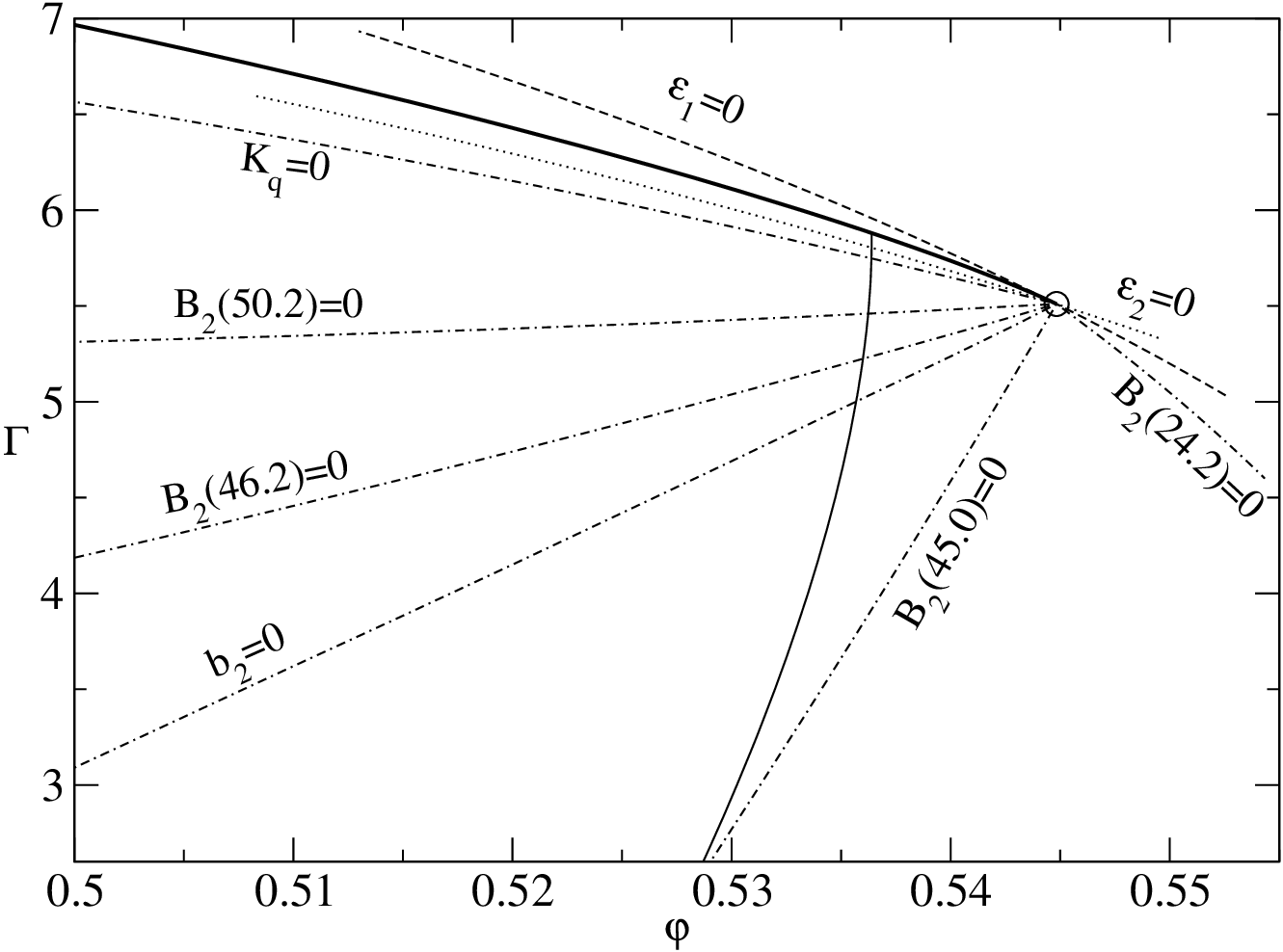}
\caption{\label{fig:swsasy:A3quadlines}Curves of vanishing quadratic 
correction in Eq.\ (\ref{eq:asy:G1G2q}) for the $A_3$-singularity
($\bigcirc$) at $\delta=0.03$. 
The $\delta=0.03$ cut through the glass-transition diagram is displayed by
full lines. The various lines are shown in the same style as in Fig.\
\ref{fig:swsasy:A4quadlines} and labeled accordingly. The line $b_2=0$,
cf. Eq.\ (\ref{eq:asy:expoxpar}), indicates the analogous line for the
MSD, cf. inset of Fig.\ \ref{fig:swsasy:MSDx}.
}
\end{figure}

Figure~\ref{fig:swsasy:A3quadlines} shows the analog of Fig.\
\ref{fig:swsasy:A4quadlines} for a cut through the glass-transition
diagram at $\delta=0.03$. The lines $\varepsilon_1=0$ and
$\varepsilon_2=0$ for the $A_3$-singularity are obtained from a smooth
transformation of the corresponding lines at the $A_4$-singularity, and
they appear in similar locations in the diagram. The line
$\varepsilon_2=0$ is again very close to the almost horizontal line of
transitions. Just below, we find again the line where $B_2(q)=0$ when
$K_q=0$. However, this now represents $q\approx 57.5$, cf.  Fig.\
\ref{fig:swsasy:fqA3}, which is a value almost twice as large as for the
corresponding line in Fig.\ \ref{fig:swsasy:A4quadlines}. For the wave
vector $q=24.2$ we find the line, where $B_2(24.2)=0$, completely in the
glass state. Taking the same value for the correction amplitude as for
$q=24.2$ at the $A_4$, $K^*_q\approx-0.6$, we obtain $q=45.0$, cf. Fig.\
\ref{fig:swsasy:fqA3} and the line labeled accordingly in Fig.\
\ref{fig:swsasy:A3quadlines}. Since the latter line comes close to the
liquid-glass-transition line we take that as a reference and estimate the
range of wave-vectors where the quadratic corrections can be put to zero
in the liquid regime to $45\lesssim q \lesssim 70$. The lines where
$B_2(q)=0$ can be rather sensitive to $q$-variation. This is demonstrated
by the curve $B_2(46.2)=0$. Although the change in the wave vector is
relatively small in comparison to $q=45.0$, the values for $K_q$ differ by
more than $20\%$ for fixed $q$ and induce a rotation of the line
$B_2(q)=0$ by quite a significant angle.

Having in mind the drastic changes in the lines where $B_2(q)=0$, it may
come with some surprise that the line for the MSD, where $b_2=0$, stays
rather robust and accessible in the liquid regime as seen in Fig.\
\ref{fig:swsasy:A3quadlines}. The variation in $q$ for the amplitudes is
reflected in changes of the localization lengths. For the
$A_3$-singularity at $\delta=0.03$ we get
\begin{equation}\label{eq:swsasy:A3rs}
r_s^\circ=0.0243\,,\quad h^\circ_\text{MSD}=0.00136\,.
\end{equation}
From Eq.\ (\ref{eq:swsasy:A4rs}) one gets $r_s^*/r_s^\circ= 1.75$ and the
square of the latter ratio, $r_s^{*\,2}/r_s^{\circ\,2}\approx 3$, is the
same as $h^*_\text{MSD}/h^\circ_\text{MSD}$. Since only the fraction
$h_\text{MSD}/r_s^2$ could introduce larger modifications in Eq.\
(\ref{eq:asy:expoxexp}), the changes in $b_2$ cancel approximately and the
line specified by $b_2=0$ experiences only minor deformations when
$\delta$ is varied. The wave vector for which the lines $B_2(q)=0$ and
$b_2=0$ are closest to each other, is $q=45.8$ at the $A_3$-singularity
for $\delta=0.03$.

\begin{figure}[htb]
\includegraphics[width=\columnwidth]{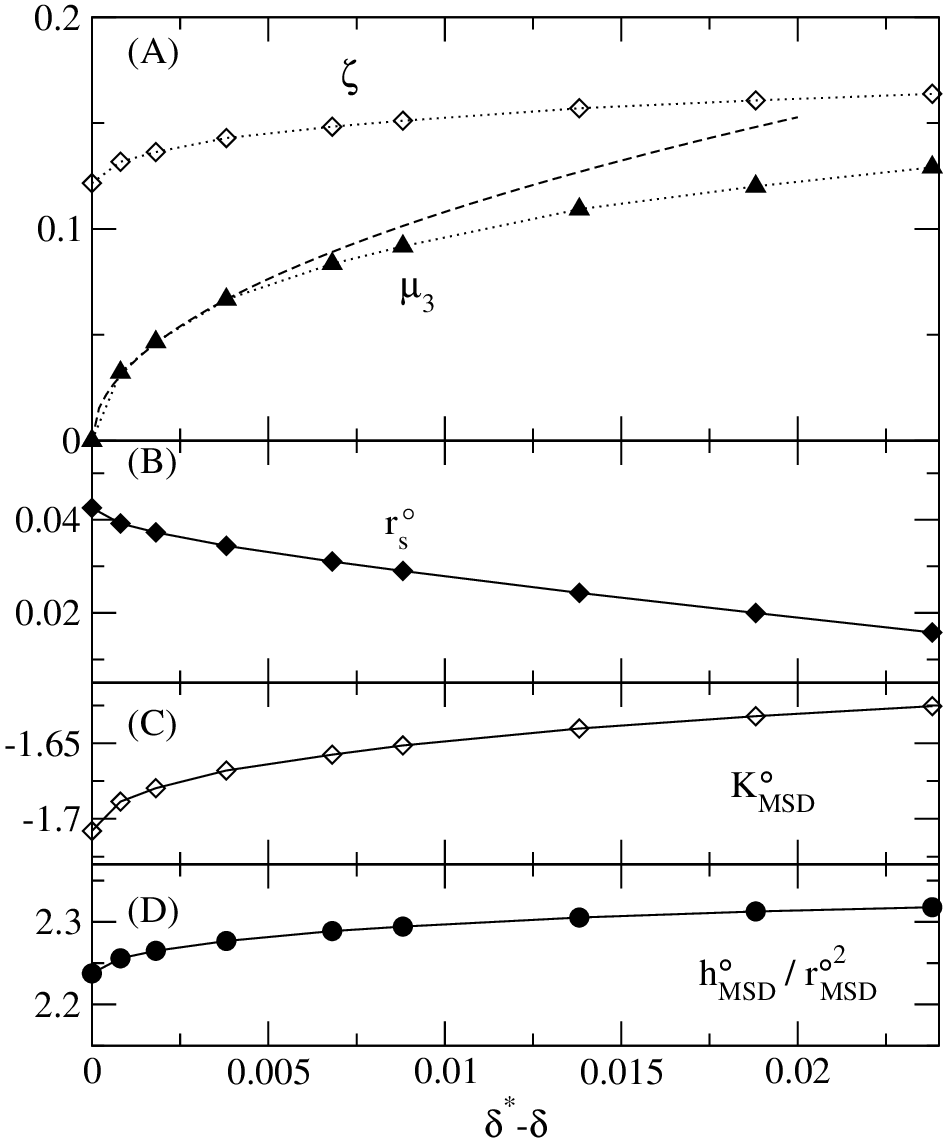}
\caption{\label{fig:swsasy:delta_var}Parameters for the asymptotic 
description at the $A_3$-singularities of the SWS for varying $\delta$.
Panel~A displays $\mu_3$ ($\blacktriangle$), Eq.\ (\ref{eq:asy:mu3}), and
$\zeta$ ($\Diamond$), Eq.\ (\ref{eq:asy:zeta}). The dashed curve shows the
asymptotic $\sqrt{\delta^*-\delta}$-law for the $\mu_3$. The localization
length $r_s^\circ$ is shown in panel~B. The correction amplitudes
$K_\text{MSD}^\circ$ and the ratios
$h_\text{MSD}^\circ/r_\text{MSD}^{\circ\,2}$ are shown in panels~C and~D.
}
\end{figure}

To corroborate the finding for the MSD from the preceding paragraph, the
parameters for the asymptotic description of the MSD at the
$A_3$-singularities are shown in Fig.\ \ref{fig:swsasy:delta_var}. The
$\mu_3$ vanish when we approach the $A_4$-singularity. The decrease close
to $\delta^*$ is described asymptotically by a square-root variation,
$\mu_3\propto \sqrt{\delta^*-\delta}$, shown by the dashed line
\cite{Sperl2003}. The smallness of $\mu_3$ indicates that all the
$A_3$-singularities are already influenced by the proximity of the
close-by $A_4$-singularity. One can take advantage of this finding and
conclude that the terms proportional to $\mu_3$ in Eq.\
(\ref{eq:asy:G1G2q}) are small. Moreover, one may neglect $B_3$ and $B_4$
in Eq.\ (\ref{eq:asy:Bcoeffs_B3}) entirely without introducing large
additional errors. The leading correction to the logarithmic decay laws is
then only quadratic also for the $A_3$-singularities.  Parameter $\zeta$
varies regularly around a finite value at $\delta^*$ but shares the
variation of $\mu_3$ at $\delta^*$ due to Eq.\ (\ref{eq:asy:mu3}). Panel~B
shows the decrease of the localization length at the $A_3$-singularity
when $\delta$ is reduced. A change of $40\%$ in $r^\circ_s$ from
$\delta=\delta^*$ to $\delta=0.03$, cf. Eqs.~(\ref{eq:swsasy:A4rs})  
and~(\ref{eq:swsasy:A3rs}), is reflected in the broadening of the
distributions in $q$ seen in Figs.\ \ref{fig:swsasy:fqA4}
and~\ref{fig:swsasy:fqA3}. This broadening is responsible for the large
variation in $q$ when comparing Fig.\ \ref{fig:swsasy:A4quadlines} with
Fig.\ \ref{fig:swsasy:A3quadlines}. It was noted in the discussion of the
inset of Fig.\ \ref{fig:swsasy:fqA3} that $K_q$ introduces the strongest
corrections for the correlation functions at the $A_4$-singularity. This
is also true for the MSD as seen in panel~C for $K_\text{MSD}$ which is
largest in absolute value at the $A_4$-singularity. The variation in
$K_\text{MSD}$ with $\delta$ is however small and does not introduce
significant changes to $a_2$ in Eq.\ (\ref{eq:asy:expoxpar}). The
amplitude $h_\text{MSD}$ is the remaining parameter entering Eq.\
(\ref{eq:asy:expoxpar}) that could alter the location of the line $b_2=0$
in the glass-transition diagram. We noted above that only the ratio
$h_\text{MSD}/r_s^{c\,2}$ needs to be considered which is shown in
panel~D. From there one infers that the ratio varies only by less than
$5\%$. We can conclude that the line of power law variation for the MSD
stays in the liquid regime even when $\delta$ is changed significantly.

\section{\label{sec:HCY}Hard Core Yukawa System}

The $A_l$-singularities occurring in MCT are topologically stable, smooth
changes in the control parameters do not challenge their existence.
Therefore, the results for the SWS can be applied also to other potentials
with a short-ranged attraction. Nevertheless, the deformation of the
potential might introduce changes large enough to be relevant for the
detection of the higher-order singularities. Among several potentials the
hard core Yukawa system (HCY) was found to differ by up to $20\%$ in
certain properties at the $A_4$-singularity from the SWS
\cite{Goetze2003b}. Since other potentials differ less we use that system
as a second example for an $A_4$-singularity.

\begin{figure}[htb]
\includegraphics[width=\columnwidth]{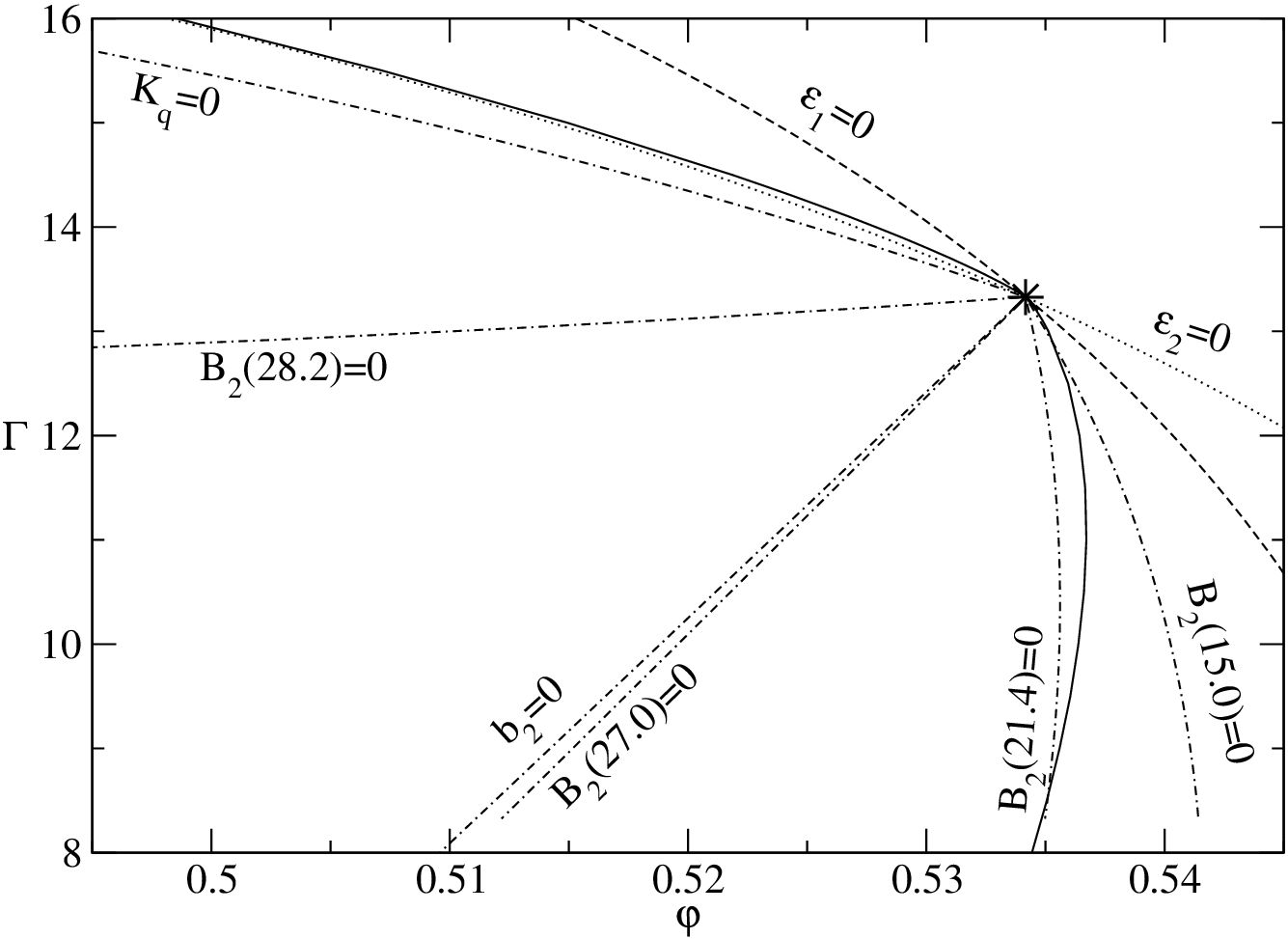}
\caption{\label{fig:HCYquadlines}Cut through the parameter space for the 
hard-core Yukawa system for $\delta=\delta^*$. 
Lines styles are the same as in Fig.\ \ref{fig:swsasy:A4quadlines}. The
wave vectors $q=15.0$, $21.4$, $27.0$, and $28.2$ are approximately
equivalent to $q=7.0$, $20.2$, $24.2$, and $27.0$ in 
Fig.\ \ref{fig:swsasy:A4quadlines}, respectively, after rescaling 
$\Gamma$ by a factor of 2.98 (see text).
}
\end{figure}

Figure~\ref{fig:HCYquadlines} shows the analog of Fig.\
\ref{fig:swsasy:A4quadlines} for the HCY. For a comparison, the
$A_4$-singularity in the SWS was mapped on top of the $A_4$-singularity in
the HCY by scaling in $\Gamma$ with a factor of $2.98$ and by a shift in 
$\varphi$ of $0.0065$.
The same transformation was applied to the lines where $B_2(q)=0$ in the
SWS. Fig.\ \ref{fig:HCYquadlines} displays the $B_2(q)=0$ lines for the
HCY that come closest to the ones shown in Fig.\ \ref{fig:swsasy:A4quadlines} 
after the mapping. The correction amplitude $K_q$ for the HCY vanishes at 
$q\approx 34$, and the range in wave vector for which $B_2(q)=0$ is lying in 
the liquid regime is shifted to higher wave vectors, $21\lesssim q\lesssim 36$ 
or $-0.9\lesssim K_q\lesssim 0.2$, in comparison to the SWS. 
For $q=27.0$ we get the line $B_2(q)=0$ for the HCY that is closest to the 
line $b_2=0$ for the MSD as compared to $B_2(24.2)=0$ in the SWS.

\section{\label{sec:conclusion}Conclusion}

Logarithmic decay or, equivalently, $1/f$~noise in the fluctuation
spectra, can arise in a number of situations and is explained by various
approaches \cite{Weissman1988}. In the log-linear representation
appropriate for the correlation functions, this decay exhibits a straight
line. To discriminate the logarithmic decay laws originating from
higher-order glass-transition singularities within MCT \cite{Goetze2002}
from other possible scenarios one needs criteria to distinguish one from
the other. The theory makes specific predictions where in the
control-parameter space the logarithmic decay is expected and how the
corrections introduce deviations from that behavior. In this paper, the
scenarios are discussed in quantitative detail for an example relevant for
studies of colloidal dynamics, the square-well system (SWS). To proceed,
specific cuts through the three-dimensional parameter space are
considered. Here, lines are identified where the corrections quadratic in
the logarithm of time vanish for a chosen wave vector $q$, cf. Fig.\
\ref{fig:swsasy:A4quadlines}. These lines emanate from the higher-order
singularity and rotate clockwise around the higher-order singularity with
increasing $q$. The correlation functions for states on these lines
exhibit decays that are linear in the logarithm of time for several orders
of magnitude in time, cf. Fig.\ \ref{fig:swsasy:A4log}. In leading order,
the slope of the decay is given by the square-root of the distance from
the higher-order singularity, Eq.\ (\ref{eq:asy:log_decay}).  The
mean-squared displacement MSD displays a power law, Eq.\
(\ref{eq:asy:expoxlaw}), that is valid on a similar line in the
control-parameter space, cf. Fig.\ \ref{fig:swsasy:MSDx}. The exponent $x$
of this subdiffusive behavior is also decreasing with the square-root of
the distance. Both the logarithmic decay and the power law are accessible
in the liquid regime. The logarithmic decay is predicted for wave vectors
$q$, which are equivalent to values of about three to four times the first
peak of the static structure factor.

In a semi-logarithmic representation for the correlation functions and a
double-logarithmic plot of the MSD, characteristic convex and concave
relaxation patterns are found when states are chosen that are off the
specified lines, cf. Figs.~\ref{fig:swsasy:A4logVvar} and
\ref{fig:swsasy:MSDV}. Due to the variation of the correction amplitude
$K_q$ in Fig.\ \ref{fig:swsasy:fqA4}, a similar variation from convex to
concave behavior is introduced by changes in the wave vector at a fixed
point in control-parameter space, cf. Fig.\ \ref{fig:swsasy:A4logqvar}.
These deviations from logarithmic behavior provide a test for the clear
identification of dynamical scenarios that are consistent with Eq.\
(\ref{eq:asy:G1G2q}) and hence originate from higher-order singularities.

When the localization at the higher-order singularity is changed by either
deforming the shape of the potential or by moving to $A_3$-singularities
at smaller ranges of the attraction, the logarithmic decay of the
correlation functions is shifted to higher wave vectors.  Whereas the
difference between the SWS and the hard-core Yukawa system at the
$A_4$-singularity is modest, cf. Fig.\ \ref{fig:swsasy:A4quadlines} and
Fig.\ \ref{fig:HCYquadlines}, the lines of vanishing quadratic correction
change drastically at the $A_3$-singularity, cf. Fig.\
\ref{fig:swsasy:A3quadlines}. In contrast, the line where the subdiffusive
power law for the MSD is valid, is robust against changes of the well
width and the potential shape, cf.  Figs.\ \ref{fig:swsasy:MSDx},
\ref{fig:swsasy:A3quadlines}, and \ref{fig:HCYquadlines}.

\begin{figure}[htb]
\includegraphics[width=\columnwidth]{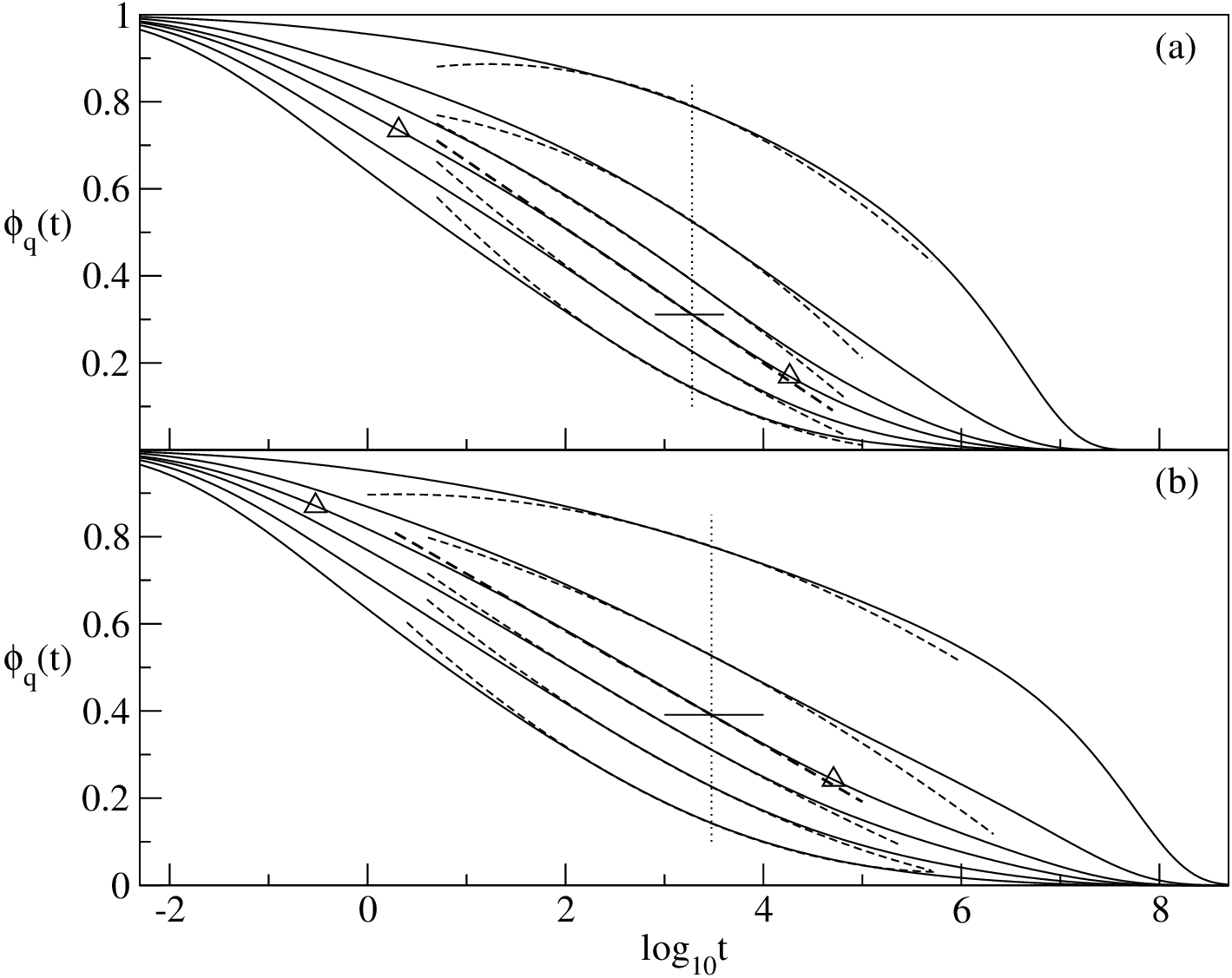}
\caption{\label{fig:conclog}Correlators for states a (panel a) and b 
(panel b) from Fig.\ \ref{fig:swsasy:A4quadlines} for wave vectors 
$q=4.2$, $20.2$, $24.2$, $27.0$, $32.2$, and $36.2$ from top to bottom. 
Full lines show the solutions of the MCT equations for the SWS, dashed
lines the approximation by Eq.\ (\ref{eq:asy:G1G2q}). Triangles mark the
$5\%$ deviation of the correlator from the approximation for $q=27.0$ and
$24.2$, respectively. The dotted vertical lines indicate the time scales
$\tau$, the short horizontal lines the corrected plateau value
$f_q+\hat{f}_q$ for $q=27.0$ and $24.2$, respectively.
}
\end{figure}

For comparing the solutions of the equations of motion, Eqs.\
(\ref{eq:mct:MCT}), (\ref{eq:mct:tagged}), and (\ref{eq:mct:MSD}), with
the asymptotic expansions, Eqs.\ (\ref{eq:asy:G1G2q}),
(\ref{eq:asy:phis}), and (\ref{eq:asy:MSD}), all parameters are calculated
explicitly except the time scale $\tau$ which is matched at the plateau.
In an experiment or a computer simulation only the correlators are
available directly. We show these in Fig.\ \ref{fig:conclog} for two
states specified in Fig.\ \ref{fig:swsasy:A4quadlines} for different wave
vectors. Since state b is closer to the $A_4$-singularity, the range of
validity for the asymptotic approximation is larger than for state a.
Especially the extension of the linear-$\log$ decay at some specific wave
vector increases when moving closer to the singularity. As noted in
connection with Fig.\ \ref{fig:swsasy:A4logqvar}, the range of validity
for the approximation by Eq.\ (\ref{eq:asy:G1G2q}) may vary with $q$.  
Partly for that reason a larger absolute curvature is attributed to the
correlators by the approximation than a fit would do. A free fit could
identify logarithmic behavior at state~b for $q=20.2$ from $t\approx 5$ to
$t\approx 5\cdot 10^7$ with a deviation of at most $5\%$. In addition,
fitting the correlator for $q=24.2$ also for $t\geqslant 10^5$ would yield
positive curvature. Therefore, a free fit in that region of the
control-parameter space tends to find the logarithmic decay at a somewhat
lower wave vector than predicted by Eq.\ (\ref{eq:asy:G1G2q}). However,
with a choice of the time scale $\tau$ that is reasonably close to the
theoretical value, the concave and convex decay patterns can still be
identified unambiguously in the correlators without invoking additional
assumptions.

\begin{figure}[htb]
\includegraphics[width=\columnwidth]{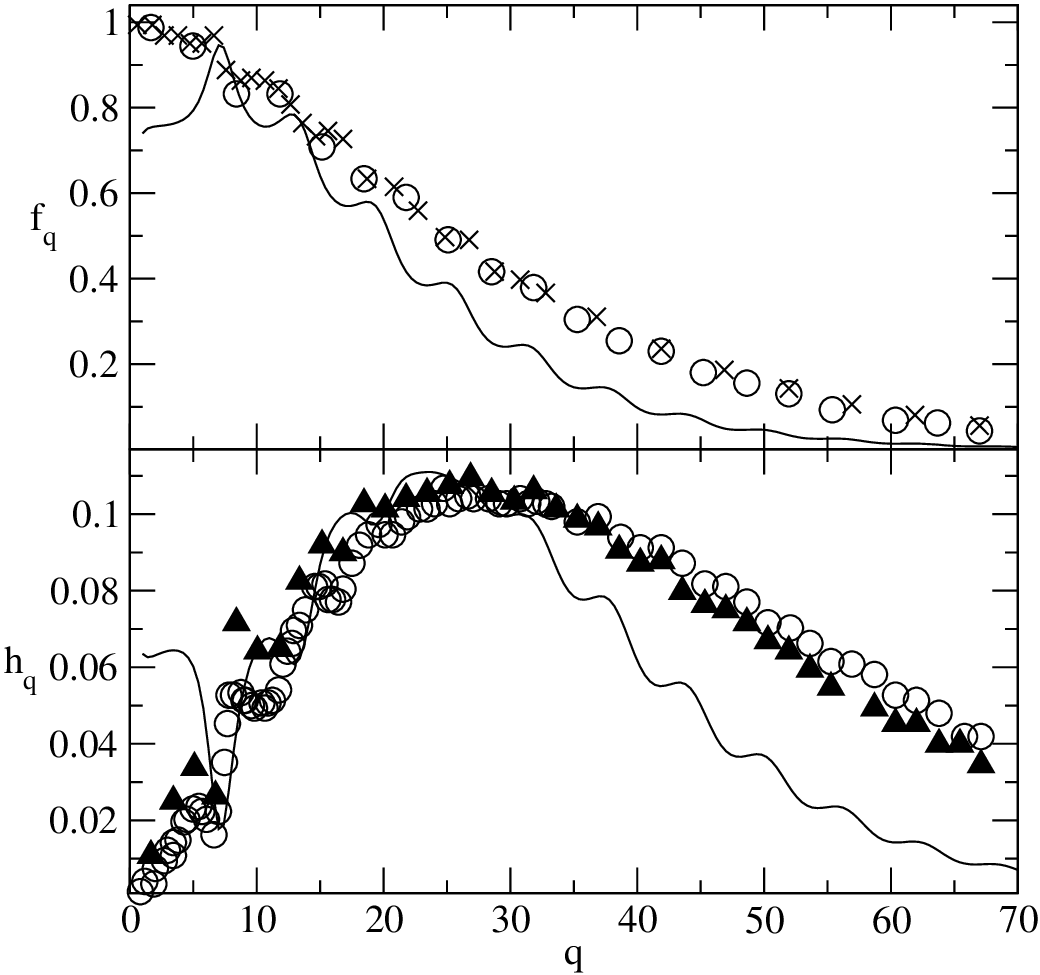}
\caption{\label{fig:concfqA4}Comparison of $f_q$ and $h_q$ from the fit to 
the simulation of two different states \cite{Sciortino2003pre} with the 
values  $f_q^*$ and $h_q^*$ for the SWS from Fig.\ \ref{fig:swsasy:fqA4}. 
For the comparison in the lower panel the theoretical values are 
multiplied by $0.14$.
}
\end{figure}

A recent molecular dynamics study of a binary mixture of square-well
particles identifies a power law with $x=0.28$ for the MSD over four
decades and a related logarithmic decay of the correlation function at a
wave vector $q=16.8$ \cite{Sciortino2003pre}. A scenario similar to Fig.\
\ref{fig:conclog} was found for the correlation functions: Upon increasing
$q$, a change from concave to convex decay is observed. For a second
state, faster decay with larger prefactors for the logarithmic decay is
reported together with a larger exponent, $x=0.44$, for the power law in
the MSD. This finding is consistent with the assumption that this second
state is further from the supposed higher-order singularity than the first
state. Different from Fig.\ \ref{fig:conclog}, in the simulation $\delta$
was changed to vary the distance while $\varphi$ and $\Gamma$ were kept
fixed. The logarithmic decay was shifted to a higher wave vector for
smaller $\delta$ \cite{Sciortino2003pre}. This is consistent with the
expectation that can be inferred from Figs.\ \ref{fig:swsasy:A4quadlines}
and \ref{fig:swsasy:A3quadlines} by observing, e.g., the rotation of the
line $B_2(24.2)=0$. The analysis of the simulation data allowed for a fit
of the values for $f_q^*$ and $h_q^*$ \cite{Sciortino2003pre}. These are
shown in Fig.\ \ref{fig:concfqA4} together with the theoretical
predictions for the SWS. The fitted parameters for both states almost fall
on top of each other for $f_q^*$. The amplitude $h_q^*$ is deduced from
the simulation data only up to some overall factor. It can be matched
reasonably by a multiplication of the theoretical prediction for $h_q^*$.
The extension in $q$ for the values obtained from MCT for the SWS are
narrower, the width at half maximum for $f_q$ differs by $15\%$. A similar
difference was observed for a binary mixture of hard spheres and agreement
between theory and simulation could be improved by using the structure
factor from the simulation as input to the MCT calculations
\cite{Foffi2003pre}. For the amplitude $h_q^*$ the locations of the maxima
disagree by $15\%$ and the width is different by $25\%$. The deviations
for $q<7$ in both $f_q$ and $h_q$ can be attributed to the effects of
mixing \cite{Foffi2003pre}.

In summary, scenarios for logarithmic decay near higher-order
glass-transition singularities are presented in this work. Some essential
predictions are supported by the results of computer simulations. This
should motivate further investigations in colloidal systems with
short-ranged attraction. In particular the power-law behavior for the MSD
including the deviations might be accessible to experiments.

\acknowledgments
I thank W.~G\"otze for valuable discussion. This work was supported by the 
Deutsche Forschungsgemeinschaft Grant Go154/13-1.

\bibliographystyle{apsrev}

\end{document}